\documentclass[pre,aps,superscriptaddress,showpacs]{revtex4-1}

\usepackage{natbib}
\usepackage{amsmath}
\usepackage{epsfig}
\usepackage{color}

\begin{document}

\title{Computing characteristic functions of quantum work in phase space }
\author{Yixiao Qian}
\author{Fei Liu}
\email[Email address: ]{feiliu@buaa.edu.cn}
\affiliation{School of Physics, Beihang University, Beijing 100191, China}

\date{\today}

\begin{abstract}
{In phase space, we analytically obtain the characteristic functions (CFs) of a forced harmonic oscillator [Talkner et al., Phys. Rev. E, 75, 050102 (2007)], a time-dependent mass and frequency harmonic oscillator [Deffner and Lutz, Phys. Rev. E, 77, 021128 (2008)], and coupled harmonic oscillators under driving forces in a simple and unified way. For general quantum systems, a numerical method that approximates the CFs to $\hbar^2$ order is proposed. {We exemplify the method with a time-dependent frequency harmonic oscillator and a family of quantum systems with time-dependent even power-law potentials.}}
\end{abstract}
\maketitle

\section{Introduction}
\label{section1}

Over the past two decades, the statistics of thermodynamic quantities in nonequilibrium quantum processes have attracted considerable interest. The major motivation of this was to extend the famous Jarzynski equality~\cite{Jarzynski1997} and Crooks equality~\cite{Crooks1999} from the classical regime to the quantum regime. In contrast to the classical thermodynamic quantities, their quantum definitions are highly challenging. For instance, although the quantum work based on the two energy measurement scheme (TEM)~\cite{Kurchan2000,Tasaki2000,Talkner2007,Campisi2011,Esposito2009,Liu2018} satisfies the Jarzynski equality and was experimentally verified~\cite{Batalhao2014,ShuomingAn2015}, there is still much debate about this definition, as it destroys the possible initial quantum coherence. Several alternatives to the TEM work were proposed in the literature~\cite{Allahverdyan2005}; see the comprehensive review by B\"{a}umer et al.\cite{Bumer2018}.

The advantages of the TEM work extend beyond satisfying the Jarzynski equality. Very recently, several studies indicated that it also follows the quantum-classical correspondence principle~\cite{Jarzynski2015,Zhu2016,Wang2017a,Garcia-Mata2017,Fei2018}. Moreover, using the phase space notion~\cite{Wigner1932}, the characteristic function (CF) (or Fourier transformation) of the distribution of the quantum work was proved to be expanded in a power series in Planck's constant $\hbar$: the zeroth order is the CF of the classical work, and only even orders have nonzero contributions to the series. Because the moments of the TEM work are equal to the derivatives of the CF with respect to the Fourier parameter at the zero point, these two properties are inherited in the moments. These results could have been expected according to the important work performed by E.P. Wigner~\cite{Wigner1932} in 1932, in which his results implied that the free energy of a quantum thermal equilibrium ensemble has the same $\hbar$-expansion properties~\cite{LandauStatistics}, while, according to the second law of thermodynamics, the mean work done on a system must be equal to or larger than the free energy change of the system.

The present paper has two aims. The first is to {show the advantages} of phase space in computing the exact CFs of several special quantum systems composed of harmonic oscillators. {Although such systems are very simple, they are good examples of exactly solved models~\cite{Husimi1953,Pedrosa1997}. A recent single ion experiment verifying the quantum Jarzynski equality is modeled as a time-dependent quantum harmonic oscillator~\cite{ShuomingAn2015}.} The second aim is to {present an approximation method} to numerically compute the CFs of general closed quantum systems. In our previous study~\cite{Fei2018}, the classical path integral formulas of the CFs approximated to $\hbar^2$ were obtained. Although these expressions have attractive forms, their numerical realizations are not the most convenient. The rest of this paper is organized as follows. In Sec.~(\ref{section2}), we briefly review {the phase space formulas} of the CFs for the TEM work. In Sec.~(\ref{section3}), we use these formulas to obtain several exact CFs of driven quantum harmonic oscillator systems. In Sec.~(\ref{section4}), for general quantum systems, an approximation method based on numerically solving partial differential equations (PDEs) in phase space is proposed. Several driven quantum models are used to exemplify the method. Section~(\ref{section5}) concludes the paper.

\section{{CFs in phase space}}
\label{section2}
We assume that the Hamiltonian of a closed quantum system is $\hat H(t)$. Throughout this paper, we use symbols with and without hats to denote quantum operators and $c$-numbers, respectively, unless otherwise stated. Given that the instantaneous energy eigenvectors and eigenvalues of the Hamiltonian are
\begin{eqnarray}
\label{eigenvectorandvaluesofclosedHamiltonian}
\hat H(t)|\varepsilon_n(t)\rangle=\varepsilon_n(t) |\varepsilon_n(t)\rangle,
\end{eqnarray}
the TEM quantum work is defined as the difference between the measured instantaneous energy eigenvalues at the end and at the beginning, that is, $W_{nm}$$=$$\varepsilon_n(t)-\varepsilon_m(0)$~\cite{Kurchan2000,Tasaki2000}. By repeating the measurement scheme many times, the probability distribution of the work can be constructed as
\begin{eqnarray}
\label{distributioninclusiveworkclosedsystem}
P(W)&=&\sum_{n,m} \delta( W-W_{nm}) P(n,t|m,0) P_m(0),
\end{eqnarray}
{where the quantum transition probability is
\begin{eqnarray}
\label{quantumtransitionprobdef}
P(n,t|m,0)=\left |\langle \varepsilon_n(t)|U(t)|\varepsilon_m(0)\rangle \right |^2,
\end{eqnarray}
}$U(t)$ is the time evolution operator of the system, and $P_m(0)$ is the probability of finding the system eigenvector $|\varepsilon_m(0)\rangle$ at time $0$. Because we initially set the quantum system in the thermal state
\begin{eqnarray}
\label{thermalstate}
{\hat \rho}_0=\frac{e^{-\beta\hat H(0)}}{{\rm Tr}[ e^{-\beta \hat H(0)} ]}
=\frac{1}{{\cal Z}_0}e^{-\beta\hat H(0)},
\end{eqnarray}
where ${\cal Z}_0$ is the partition function at time $0$ and $\beta$ is the inverse temperature, we have $P_m(0)$$=$$\exp[-\beta\varepsilon_m(0)]/{\cal Z}_0$. Rather than focusing on the distribution~(\ref{distributioninclusiveworkclosedsystem}), we devote our attention to the CF of the quantum work~\cite{Talkner2007}. A simple argument shows that the CF can be expressed as a trace over the operators~\cite{Fei2018},
\begin{eqnarray}
\label{CFworkclosedsystem}
\Phi(\eta)
&=&{\rm Tr}\left[ e^{i\eta\hat{
H}(t)}{U}(t)e^{-i\eta \hat{H}(0)}{ \hat{\rho}_0 U}^\dag(t)\right] \nonumber \\
&= &{\rm Tr}\left[ e^{i\eta\hat{H}(t)} \hat \varrho(t)\right].
\end{eqnarray}
Importantly, the operator $\hat\varrho$ in the above second equation satisfies the von Neumann equation,
\begin{eqnarray}
\label{vonNeumannequation}
{\partial_t}\hat{\varrho}(t)=\frac{1}{i\hbar}[\hat{H}(t),\hat{\varrho}(t)],
\end{eqnarray}
and its initial condition is
\begin{eqnarray}
\label{thermalinitialcondition}
\hat\varrho_0=\frac{1}{{\cal Z}_0}e^{-(i\eta+\beta) \hat H(0)},
\end{eqnarray}
or a complex extension of the thermal state~(\ref{thermalstate}). If we solve Eq.~(\ref{vonNeumannequation}) and obtain Eq.~(\ref{CFworkclosedsystem}), then the distribution of the quantum work can be calculated by performing an inverse Fourier transformation over $\Phi(\eta)$.

Let us consider an $n$-dimensional quantum system with the Hamiltonian
\begin{eqnarray}
\label{generalHamiltonian}
\hat H(t) =\sum_{i=1}^n\frac{\hat {p_i}^2}{2m_i} + U(\hat{\bf x}, t),
\end{eqnarray}
where $m_i$ is the mass of the $i$th quantum particle, $U$ is the time-dependent potential, $\hat{ \bf x}^T$$=$$(\hat x_1,\cdots,\hat x_n)$ and $\hat{ \bf p}^T$$=$$(\hat p_1,\cdots,\hat p_n)$ are the position and momentum operator vectors, respectively, and the superscript $T$ denotes the transpose. We can re-express Eqs.~(\ref{CFworkclosedsystem}) and~(\ref{vonNeumannequation}) in phase space. Given the Weyl symbol $[\hat\varrho(t)]_w$$=$$P({\bf z},t)$, where we use the subscript $w$ to denote the Weyl symbol of a quantum operator throughout this paper, the von Neumann equation~(\ref{vonNeumannequation}) in phase space is~\cite{Wigner1932,Hillery1984,Polkovnikov2010,Schleich2001}
\begin{eqnarray}
\label{vonNeumannequationinphasespace}
{\partial_t P}({\bf z},t)& =&-\frac{2}{\hbar} H({\bf z},t)\sin\left(\frac{\hbar\Lambda}{2}\right)P({\bf z},t),
\end{eqnarray}
where ${\bf z}^T$$=$$({\bf x}^T,{\bf p}^T)$ is the phase point, $H({\bf z},t)$ is the c-number Hamiltonian of Eq.~(\ref{generalHamiltonian}), $\Lambda$ is the negative Poisson bracket, i.e.,
\begin{eqnarray}
\label{Lambda}
&&\overleftarrow{\partial_{\bf p}}\overrightarrow{\partial_{\bf x}}-\overleftarrow{\partial_{\bf x}}\overrightarrow{\partial_{\bf p}}
= \sum_{i=1}^{n}\overleftarrow\partial_{p_i}\overrightarrow\partial_{x_i}-\overleftarrow\partial_{x_i}\overrightarrow\partial_{p_i},
\end{eqnarray}
and the arrows indicate the directions in which the derivatives act~\cite{Imre1967}. Correspondingly, the initial condition is given by
\begin{eqnarray}
\label{initialconditionHCOWeylsymbol}
P({\bf z},0)=\frac{1}{(2\pi\hbar)^n}\left[\hat\varrho_0\right]_w({\bf z}).
\end{eqnarray}
After solving PDE~(\ref{vonNeumannequationinphasespace}), we compute the CF of the quantum work by an integral,
\begin{eqnarray}
\label{CFworkclosedsysteminphasespace}
\Phi(\eta)=\int_{-\infty}^{+\infty} d{\bf z} \left[e^{i\eta \hat H(t)}\right]_w({\bf z})P({\bf z},t).
\end{eqnarray}

{Here we make two comments. First, if the initial state of a quantum system is not thermal but commutative with the initial Hamiltonian ${\hat H}(0)$, then Eqs.~(\ref{CFworkclosedsystem})-(\ref{CFworkclosedsysteminphasespace}) are still applicable except that the right-hand side of Eq.~(\ref{thermalinitialcondition}) is changed to $\exp[-i\eta{\hat H}(0)]\hat\rho_0$. We will show such a case in Sec.~\ref{section3D}. Second, based on Eqs.~(\ref{vonNeumannequationinphasespace}) and~(\ref{CFworkclosedsysteminphasespace}), the proof of the Jarzynski equality is straightforward: let $\eta=i\beta$; the solution of Eq.~(\ref{vonNeumannequationinphasespace}) is $P({\bf z},t)$$=$$1/(2\pi\hbar)^n{\cal Z}_0$, and then,
\begin{eqnarray}
\label{JE}
\Phi(i\beta)=\left\langle e^{-\beta W}\right\rangle =\frac{{\cal Z}_t}{{\cal Z}_0},
\end{eqnarray}
where ${\cal Z}_t$ is the instantaneous partition function of the quantum system at time $t$, i.e.,
\begin{eqnarray}
{\cal Z}_t
=\frac{1}{(2\pi\hbar)^n}\int_{-\infty}^{+\infty} d{\bf z} \left[e^{i\eta \hat H(t)}\right]_w.
\end{eqnarray}}

\section{Exact CFs of driven quantum harmonic oscillators }
\label{section3}
{Due to the difficulties in obtaining the Weyl symbol of the exponential Hamiltonian operator in Eq.~(\ref{CFworkclosedsysteminphasespace}) and solving the complicated PDE~(\ref{vonNeumannequationinphasespace}), we do not think that computing the CFs of the quantum work would be easier in phase space than in other quantum representations.} However, for the quantum systems composed of harmonic oscillators, phase space indeed exhibits some advantages due to the absence of these two difficulties. Expanding Eq.~(\ref{vonNeumannequationinphasespace}) in powers of $\hbar$, when the power of the harmonic potentials equals two, Eq.~(\ref{vonNeumannequationinphasespace}) immediately reduces to the familiar Liouville equation for the classical harmonic oscillators~\cite{Wigner1932}. Hence, Eq.~(\ref{CFworkclosedsysteminphasespace}) is simplified to
\begin{eqnarray}
\label{CFworkclosedsysteminphasespaceoscillator}
\Phi(\eta)=\frac{1}{(2\pi\hbar)^n {\cal Z}_0} \int_{-\infty}^{+\infty} d{\bf z}_0 \left[e^{i\eta \hat H(t)}\right]_w({\bf z}_t)\left[e^{-(i\eta+\beta) \hat H(0)}\right]_w({\bf z}_0).
\end{eqnarray}
${\bf z}_t$ therein is the dynamic solution of the special classical system at time $t$, and the system starts from the initial phase point ${\bf z}_0$. To arrive at Eq.~(\ref{CFworkclosedsysteminphasespaceoscillator}), we have applied the Liouville theorem and initial condition~(\ref{thermalinitialcondition}). It is important to emphasize that the components of ${\bf z}_t$ are time-dependent linear combinations of the components of ${\bf z}_0$ since the dynamics of the classical harmonic oscillators are linear. Moreover, the key Weyl symbol of the exponential Hamiltonian of the harmonic oscillator was already known in very early work~\cite{Imre1967,Hillery1984}. For instance, given a simple harmonic oscillator in one dimension,
\begin{eqnarray}
\label{standardoscillatorHoperator}
\hat H_s=\frac{\hat p^2}{2m}+\frac{m\omega^2}{2}\hat x^2,
\end{eqnarray}
where $m$ is the mass of the particle and $\omega$ is the angular frequency,
\begin{eqnarray}
\label{WeylexponentialHamiltonianoscillator}
\left[e^{i\eta \hat H_s}\right]_w(z)= {\rm sech}\left(\frac{i\eta\hbar\omega}{2}\right)\exp\left[\frac{2}{\hbar\omega}{\rm tanh}\left(\frac{i\eta\hbar\omega }{2}\right)H(z)\right].
\end{eqnarray}
{We use the symbol $z$ rather than the bold font $\bf z$ to denote the phase point of special one-dimensional systems.} In the remainder of this section, we apply Eqs.~(\ref{CFworkclosedsysteminphasespaceoscillator}) and~(\ref{WeylexponentialHamiltonianoscillator}) to compute the CFs of three quantum systems composed of harmonic oscillators.

\subsection{Oscillator driven by a time-dependent force}
The first system is a one-dimensional quantum harmonic oscillator driven by a time-dependent force, for which the Hamiltonian is given by~\cite{Talkner2008}
\begin{eqnarray}
\label{singleoscillatorhamiltonian}
\hat{H}_F(t)=\frac{\hat p^2}{2m}+\frac{m\omega^2\hat x^2}{2} - F(t)\hat x,
\end{eqnarray}
where $F(t)$ is the external driving force and is set to zero at time 0. We obtain the Weyl symbol of the exponential Eq.~(\ref{singleoscillatorhamiltonian}) by slightly modifying Eq.~(\ref{WeylexponentialHamiltonianoscillator}) to
\begin{eqnarray}
\label{WignerOmega}
\left[e^{i\eta \hat H_F(t)}\right]_w(z)= \exp\left[\frac{-i\eta F^2(t)}{2m\omega^2} \right ]{\rm sech}\left(\frac{i\eta\hbar\omega}{2}\right)\exp\left[\frac{2}{\hbar\omega}{\rm tanh}\left(\frac{i\eta\hbar\omega }{2}\right)H_F'(z,t)\right],
\end{eqnarray}
where
\begin{eqnarray}
H_F'(z,t)=\frac{p^2}{2m}+\frac{1}{2} m\omega^2 \left[ x -\frac{F(t)}{m\omega^2}\right ]^2,
\end{eqnarray}
or the Hamiltonian of the classical oscillator with an instantaneous equilibrium position $F(t)/m\omega^2$. Substituting this equation and the dynamic solution of the classical oscillator, Eqs.~(\ref{dynamicsolutiondrivingoscillator1}) and~(\ref{dynamicsolutiondrivingoscillator2}), into Eq.~(\ref{CFworkclosedsysteminphasespaceoscillator}) and noting that
\begin{eqnarray}
{\cal Z}_0=\frac{1}{2}{\rm sinh}^{-1}\left(\frac{\beta\hbar\omega}{2}\right),
\end{eqnarray}
we obtain the exact CF of the quantum system by performing a simple Gaussian integration,
\begin{eqnarray}
\label{CFharmonicoscillatortimedependentforce}
\Phi(\eta)&=& \exp\left[-\frac{i\eta F^2(t) }{2m\omega^2}\right]\exp\left[\frac{2{\cal W}(t)/\hbar\omega}{\coth(i\hbar\omega\eta/2)-\coth(i\hbar\omega\eta/2+\hbar\omega\beta/2)} \right],
\end{eqnarray}
where
\begin{eqnarray}
\label{classicalwork}
{\cal W}(t)=\frac{1}{2m\omega^2}\left| \int_0^t ds \dot{F}(s) e^{i\omega s}\right |^2,
\end{eqnarray}
and the dot denotes a derivative with respect to time. {Note that the function ${\cal W}(t)$ is the classical work done on classical system $H_F'(z)$ if the oscillator is initially at rest. For further details, see Appendix A. The simplification of Eq.~(\ref{CFharmonicoscillatortimedependentforce}) is the same as that obtained by Talkner et al.~\cite{Talkner2008}; see Eq.~(\ref{Talknerequation}). They applied the first equation of~(\ref{CFworkclosedsystem}) and solved the diagonal matrix element of the exponential Hamiltonian operator in the Heisenberg picture. Eq.~(\ref{CFharmonicoscillatortimedependentforce}) clearly illustrates that the quantum work of system~(\ref{singleoscillatorhamiltonian}) is always equal to a sum of multiples of $\hbar\omega$ and $-F^2(t)/2m\omega^2$: the former is seen by expanding the second exponential function into a Laurent series of $\exp(i\eta\hbar\omega)$, and the latter comes from the inverse Fourier transformation of the first exponential function. Talkner et al. solved the corresponding quantum work distribution by performing an inverse Fourier transformation over the CF~(\ref{CFharmonicoscillatortimedependentforce}). }

\subsection{Oscillator with time-dependent mass and frequency}
Deffner and Lutz~\cite{Deffner2008} studied the quantum work distribution of a harmonic oscillator with time-varying frequency. Here, we extend their model to the case with both time-dependent mass and frequency. The Hamiltonian is
\begin{eqnarray}
\label{Hamiltoniantimemassfrequency}
\hat H_{P}(t)=\frac{{\hat p}^2}{2m_t}+\frac{1}{2}m_t{\omega^2_t}{\hat x}^2,
\end{eqnarray}
where the subscript $t$ denotes the time dependence of the parameters. Although we generally do not know the exact dynamics of the classical oscillator with time-dependent mass and frequency, since the system is still linear, its formal solution is given by
\begin{eqnarray}
\label{formalsolution1}
x_t&=&X(t)x_0+ Y(t)p_0, \\
\label{formalsolution2}
p_t&=&Q(t)x_0+R(t)p_0,
\end{eqnarray}
where $(X(t),Q(t))$ and $(Y(t),R(t))$ are the formal solutions of the classical system satisfying the specific initial phase points $(1,0)$ and $(0,1)$, respectively. Substituting these and the Weyl symbol~(\ref{WeylexponentialHamiltonianoscillator})~\footnote{Now its mass and frequency have time parameters.} into Eq.~(\ref{CFworkclosedsysteminphasespaceoscillator}) and again performing a Gaussian integration, we obtain
\begin{eqnarray}
\label{CFtimevaryingmassandfrequency}
\Phi(\eta)=\frac{\sqrt{2}\left(1-e^{-\beta\hbar\omega_0}\right)e^{i\eta\hbar(\omega_t- \omega_0)/2}}
{[\Sigma(t)(1-e^{2i\eta\hbar\omega_t})(1-e^{-2\left(i\eta+\beta\right)\hbar\omega_0})+
(1+e^{2i\eta\hbar\omega_t})(1+e^{-2\left(i\eta+\beta\right)\hbar\omega_0})-4e^{i\eta\hbar\omega_t}e^{-\left(i\eta+\beta\right)\hbar\omega_0}]^{1/2}},
\end{eqnarray}
where
\begin{eqnarray}\label{Bigsigma}
\Sigma(t)=\frac{1}{2} \left[ m_0 \omega_0 \frac{H_P(Y(t),R(t))}{\omega_t}+\frac{1}{m_0\omega_0} \frac{H_P(X(t),Q(t))}{\omega_t} \right].
\end{eqnarray}
See Appendix A for more details.

{Eq.~(\ref{CFtimevaryingmassandfrequency}) is long. To check its correctness, we test whether it satisfies the Jarzynski equality~(\ref{JE}) given $\eta=i\beta$ therein and find that
\begin{eqnarray}
\label{CFforJEvaryingmassfreqoscillator}
\Phi(i\beta)=\frac{\sinh\left({\beta\hbar\omega_0}/{2}\right)}{\sinh\left({\beta\hbar\omega_t}/{2}\right)}.
\end{eqnarray}
Because the instantaneous partition functions of quantum system~(\ref{Hamiltoniantimemassfrequency}) are ${\cal Z}_t$$=$$1/2\sinh(\beta\hbar\omega_t/2)$, the right-hand side of Eq.~(\ref{CFforJEvaryingmassfreqoscillator}) is indeed the ratio of ${\cal Z}_t$ over ${\cal Z}_0$. We also note that different realizations of $m_t$ and $\omega_t$ only influence CF~(\ref{CFtimevaryingmassandfrequency}) through the function $\Sigma(t)$. This function, in particular, measures the degree of adiabaticity: if the mass and frequency are changed infinitely slowly, then the two ratios of the Hamiltonian to the frequency in Eq.~(\ref{Bigsigma}) are the adiabatic invariants~\cite{Landau1976}, and then, $\Sigma(t)$$=$$1$. These two observations were made by Deffner and Lutz~\cite{Deffner2008} when they considered a quantum harmonic oscillator with a constant mass and time-dependent frequency.
}
\subsection{Coupled oscillators under driving forces}
The last system consists of multi-harmonic oscillators. The simplest model is a driving force system composed of two identical but distinguishable harmonic oscillators. Its Hamiltonian is given by
\begin{eqnarray}
\label{twooscillatorhamiltonian}
\hat{H}(t)&=&\frac{\hat p_1^2}{2m}+\frac{m\omega^2\hat x_1^2}{2} -F(t)\hat x_1 +\frac{1}{2}k (\hat x_1-\hat x_2)^2 +\frac{\hat p_2^2}{2m}+\frac{m\omega^2\hat x_2^2}{2}.
\end{eqnarray}
Here, we assume that only one of the oscillators is driven by a time-dependent force, $F(t)$. At time 0, $F(0)$ is also set to zero, and these two coupled oscillators are at thermal equilibrium with a heat bath with an inverse temperature $\beta$. According to the well-known strategy for managing coupled oscillators~\cite{goldstein1980}, we attempt to simultaneously diagonalize both the kinetic and potential energy terms of Eq.~(\ref{twooscillatorhamiltonian}) by applying a canonical transformation:
\begin{eqnarray}
\label{canonicaltransformation}
\hat{\bf x}={\bf A}\hat{\bf Q}, \hspace{0.5cm}\hat{\bf p}={\bf MA}\hat{\bf P},
\end{eqnarray}
where the mass matrix and transformation matrix are
\begin{eqnarray}
{\bf M}=\left[
\begin{array}{cc}
m & 0 \\
0 & m \\
\end{array}
\right],\hspace{0.5cm}
{\bf A}=\frac{1}{\sqrt{2m}}\left [
\begin{array}{cc}
1 & 1 \\
-1 & 1 \\
\end{array}
\right ],
\end{eqnarray}
and the new phase coordinates are $\hat {\bf Q}^T$$=$$(\hat Q_1,\hat Q_2)$ and $\hat {\bf P}^T$$=$$(\hat P_1,\hat P_2)$. Under the transformation, Eq.~(\ref{twooscillatorhamiltonian}) is equal to $\sum_{i=1}^2 \hat H_i(t)$, or the sum of the Hamiltonian of modes $1$ and $2$, where
\begin{eqnarray}
\label{Hindependentoscillators}
\hat H_i(t)&=&\frac{\hat P_i^2}{2}+\frac{\omega_i^2\hat Q_i^2}{2} - \frac{F(t)}{\sqrt{2m}}\hat Q_i,
\end{eqnarray}
and the normal frequencies are $\omega_1^2$$=$$\omega^2+2k/m$ and $\omega_2^2=\omega^2$. Because the Weyl symbol of exponential Eq.~(\ref{twooscillatorhamiltonian}) is equal to the product of two independent Weyl symbols of the exponential Hamiltonian $\hat H_i(t)$ and the classical dynamics of the two coupled oscillator are reduced to the dynamics of two independent modes, CF~(\ref{CFworkclosedsysteminphasespaceoscillator}) of the two coupled quantum oscillators is given by
\begin{eqnarray}
\label{CFworkclosedsysteminphasespaceoscillator2}
\Phi(\eta)&=& \prod_{i=1}^2 \frac{1}{2\pi\hbar {\cal Z}_i } \int_{-\infty}^{+\infty} d{{\bf Z}_i}_0\left[e^{i\eta \hat H_i(t)}\right]_w({{\bf Z}_i}_t)\left[e^{-(i\eta+\beta) \hat H_i(0)}\right]_w({{\bf Z}_i}_0)\nonumber \\
&=&\prod_{i=1}^2\Phi_i(\eta),
\end{eqnarray}
where ${\bf Z}_i^T$$=$$({ Q}_i,{P}_i)$ is the phase point of the independent oscillator, and $\Phi_i(\eta)$ is the CF~(\ref{CFharmonicoscillatortimedependentforce}) of the quantized modes with the specified Hamiltonian~(\ref{Hindependentoscillators}). To arrive at the above result, we have used the fact that under the transformation ${\bf z}^T$$\rightarrow$${\bf Z}^T$$=$$(Q_1,Q_2,P_1,P_2)$, the Jacobian determinant $|d{\bf z}_0/d{\bf Z}_0|$ equals $1$. Of course, we can also easily prove Eq.~(\ref{CFworkclosedsysteminphasespaceoscillator2}) according to the original definition of the CF, Eq.~(\ref{CFworkclosedsystem}), by applying the same canonical transformation~(\ref{canonicaltransformation}). {Because the inverse Fourier transform of Eq.~(\ref{CFworkclosedsysteminphasespaceoscillator2}) is a convolution of two quantum work probability distributions of the respective CFs, this result implies that the quantum work done on the coupled oscillators is equal to a sum of two components acting on the two independent modes. To be more specific, for a general case where the ratio $\omega_1/\omega_2$ is irrational, the distribution of the quantum work for the coupled oscillators is
\begin{eqnarray}
\label{probworkcoupledoscillators}
P(W)=\sum_{\bf q} \delta(W-W_{\bf q})\prod_{i=1}^2 P_{q_i},
\end{eqnarray}
where the quantum work
\begin{eqnarray}
W_{\bf q }=\sum_{i=1}^2  W_{q_i}=\sum_{i=1}^2 \left[q_i\hbar\omega_i -\frac{1}{2\omega_i^2} \left(\frac{F(t)}{\sqrt{2m}}\right)^2 \right],
\end{eqnarray}
$\bf q$$=$$(q_1,q_2)$, and $q_i$ ($i$$=$$1,2$) are arbitrary integers. In Eq.~(\ref{probworkcoupledoscillators}), $P_{q_i}$ is the probability of the quantum work $W_{q_i}$ acting on the mode $i$~\cite{Talkner2008}:
\begin{eqnarray}
P_{q_i}=\exp\left[ -\frac{{\cal W}_i}{\hbar\omega_i}\coth\left(\frac{\hbar\omega_i\beta}{2}\right)   \right]\exp\left(\frac{\hbar\omega_i\beta q_i}{2} \right) I_{q_i}\left[\frac{{\cal W}_i/\hbar\omega_i}{\sinh(\hbar\omega_i\beta/2)}  \right],
\end{eqnarray}
where $I_{q_i}(x)$ is the modified Bessel function of the fist kind of order $q_i$, and ${\cal W}_i$ is the classical work~(\ref{classicalwork}) of force $F(t)/\sqrt{2m}$ acting on the same mode.

}

The Hamiltonian of the two coupled identical harmonic oscillators is a special case. Its extension to the general situation with arbitrary masses and coupling constants is straightforward, e.g.,
\begin{eqnarray}
\label{multiharmonicoscillators}
\hat H(t)=\frac{1}{2} \hat {\bf p}^T {\bf M}^{-1} \hat {\bf p} - {\bf F}^T(t)\hat {\bf x}+\frac{1}{2} \hat {\bf x}^T {\bf K} \hat {\bf x},
\end{eqnarray}
where $\bf M$ and $\bf K$ are the symmetric mass matrix and coupling constant matrix, respectively, and ${ \bf F}^T(t)$$=$$(F_1(t),\cdots,F_n(t))$ are the possible time-dependent forces exerted on each oscillator. This extension shall be significant, e.g., in the study of quantum Brownian motion~\cite{Ken2018a,Ken2018b}. The mechanical theorem ensures the existence of an orthogonal transformation matrix ${\bf A}$ for which
\begin{eqnarray}
{\bf A}^{T} {\bf M} {\bf A}&=&{\bf I}, \nonumber \\
{\bf A}^T {\bf K} {\bf A}&=&{\bf \Omega}={\rm diag}(\omega_1,\cdots,\omega_n),
\end{eqnarray}
where ${\bf I}$ is the $n$-dimensional identity matrix, and $\omega_i$ ($i$$=$$1,\cdots,n$) are the normal frequencies~\cite{goldstein1980}. The transformation decouples the general Hamiltonian~(\ref{multiharmonicoscillators}) into a sum of independent modes with $\hat H_i$ analogous to Eq.~(\ref{Hindependentoscillators}) except for the simple replacement
\begin{eqnarray}
\frac{F(t)}{\sqrt{2m}}\rightarrow {\cal F}_i(t)=\sum_{j=1}^nF_j(t)({\bf A})_{ji}.
\end{eqnarray}
Applying the same argument, we find that the CF of the multi-coupled quantum harmonic oscillators is equal to a multiplication of $n$ CFs of the independent quantized modes: the masses of these modes are the same and equal to $1$, their frequencies are the normal frequencies $\omega_i$, and they are subjected to the driving forces ${\cal F}_i(t)$. Eq.~(\ref{probworkcoupledoscillators}) has a similar extension.

{

\subsection{CF with a microcanonical initial state}
\label{section3D}
All initial states of the previous three models are thermal states. Here, we compute the CF of the quantum harmonic oscillator~(\ref{singleoscillatorhamiltonian}) with the microcanonical initial state $|\varepsilon_m(0)\rangle$ in phase space~\cite{Talkner2008}. As we mentioned in Sec.~\ref{section2}, Eq.~(\ref{CFworkclosedsysteminphasespaceoscillator}) is now modified to
\begin{eqnarray}
\label{CFworkclosedsysteminphasespaceoscillatormicrocanoical}
\Phi_m(\eta)&=&\frac{1}{2\pi\hbar} \int_{-\infty}^{+\infty} d{ z}_0 \left[e^{i\eta \hat H_F(t)}\right]_w({z}_t)\left[e^{-i\eta \hat H_F(0)}| \varepsilon_m(0) \rangle\langle \varepsilon_m(0)|\right]_w({z}_0).
\end{eqnarray}
For the specific quantum oscillator, the Weyl symbol of its instantaneous eigenvector $|\varepsilon_m(t)\rangle$ is known to be~\cite{Hillery1984}
\begin{eqnarray}
\label{Weylsymbolntheigenvector}
\left[|\varepsilon_m(t)\rangle\langle \varepsilon_m(t) \right]_w(z)=2(-1)^m \exp\left[-\frac{2}{\hbar\omega} H_F'(z,t)\right] L_m \left[ \frac{4}{\hbar\omega} H_F'(z,t)\right ],
\end{eqnarray}
where $L_m$ denotes the $m$th Laguerre polynomial. Substituting Eqs.~(\ref{WignerOmega}) and~(\ref{Weylsymbolntheigenvector}) into Eq.~(\ref{CFworkclosedsysteminphasespaceoscillatormicrocanoical}) and using the generating function of the Laguerre polynomial, we obtain ~\cite{Talkner2008}
\begin{eqnarray}
\label{CFmicrocanonicalinitialstate}
\Phi_m(\eta)=\exp\left[-\frac{i\eta F(t)^2}{2m\omega^2}\right] \exp\left[\frac{{\cal W}(t)}{\hbar\omega}\left(e^{i\eta\hbar\omega}-1\right)\right]
L_m\left[ \frac{4{\cal W}(t)}{\hbar\omega}\sin^2\left(\frac{\eta\hbar\omega}{2}\right) \right].
\end{eqnarray}
Appendix A provides some details.

Based on Eq.~(\ref{CFmicrocanonicalinitialstate}), Talkner et al. obtained the corresponding probability distribution of the quantum work by performing an inverse Fourier transformation. However, for this relatively simple quantum system, its work distribution with the microcanonical initial state $|\varepsilon_m(0)\rangle$,
\begin{eqnarray}
\label{workdefinitionmicrocanonical}
P(W|m,0)&=&\sum_{n}\delta( W-W_{nm}) P(n,t|m,0),
\end{eqnarray}
can be obtained by directly computing its quantum transition probability~\cite{Husimi1953,Campisi2008}. Here, we are not ready to repeat previous derivations but point out an unnoticed fact that can be clearly observed in phase space. To simplify notations, we use the natural units in the following, $\hbar$$=$$\omega$$=$$m$$=$$1$. In phase space, the quantum transition probability~(\ref{quantumtransitionprobdef}) of the one-dimensional quantum harmonic oscillators is
\begin{eqnarray}
\label{distributioninclusiveworkmicrocanonical}
P(n,t|m,0)
&=&\frac{1}{2\pi\hbar}\int_{-\infty}^{+\infty} d{z_t} \left[|\varepsilon_n(t)\rangle\langle \varepsilon_n(t) \right]_w( z_t)\left[|\varepsilon_m(0)\rangle\langle \varepsilon_m(0) \right]_w({z}_0).
\end{eqnarray}
The reason is analogous to that we considered in obtaining Eq.~(\ref{CFworkclosedsysteminphasespaceoscillator}). We rewrite Eq.~(\ref{Weylsymbolntheigenvector}) as
\begin{eqnarray}
\label{Weylsymbolntheigenvectornatural1}
\left[|\varepsilon_n(t)\rangle\langle \varepsilon_n(t) \right]_w({ z_t} )=2(-1)^n \exp\left[-2 H_s(z_t-z_n(t))\right] L_n \left[ 4H_s(z_t-z_n(t))\right ]
\end{eqnarray}
by using the $c$-number version of the simple Hamiltonian~(\ref{standardoscillatorHoperator}), where $z_n(t)$$=$$(f(t),0)$ is the equilibrium phase point of $H_F'(z,t)$, $f(t)$ is the dimensionless form of force $F(t)$, and the eigenvalue $\varepsilon_n(t)$ is $(n+1/2)-f(t)^2/2$ in the natural units. Before performing an integration, we note that Eq.~(\ref{Weylsymbolntheigenvectornatural1}) is rotationally symmetric about $z_n(t)$. Because of the dynamic solution Eq.~(\ref{vectorformofdynamicalsolution}), the other Weyl symbol is
\begin{eqnarray}
\label{Weylsymbolntheigenvectornatural2}
\left[|\varepsilon_m(0)\rangle\langle \varepsilon_m(0) \right]_w({z}_0(z_t))=2(-1)^m \exp\left[-2 H_s(z_t -D(t) )\right] L_n \left[ 4H_s(z_t -D(t) )\right ],
\end{eqnarray}
where $D(t)$$=$$(l(t),\dot{l}(t))$ and the function $l(t)$ is defined in Eq.~(\ref{definitionoflfunction}). The explanation of Eq.~(\ref{Weylsymbolntheigenvectornatural2}) is that, according to the Liouville equation of the force-driven harmonic oscillator, the evolution of the initial Weyl symbol, $\left[|\varepsilon_m(0)\rangle\langle \varepsilon_m(0) \right]_w(z)$, which is initially rotationally symmetric about the origin, is simply rotated and translated to the new phase point $D(t)$. Fig.~(\ref{fig1}) schematically depicts these two Weyl symbols in phase space.
\begin{figure}
\includegraphics[angle=0,width=1\linewidth]{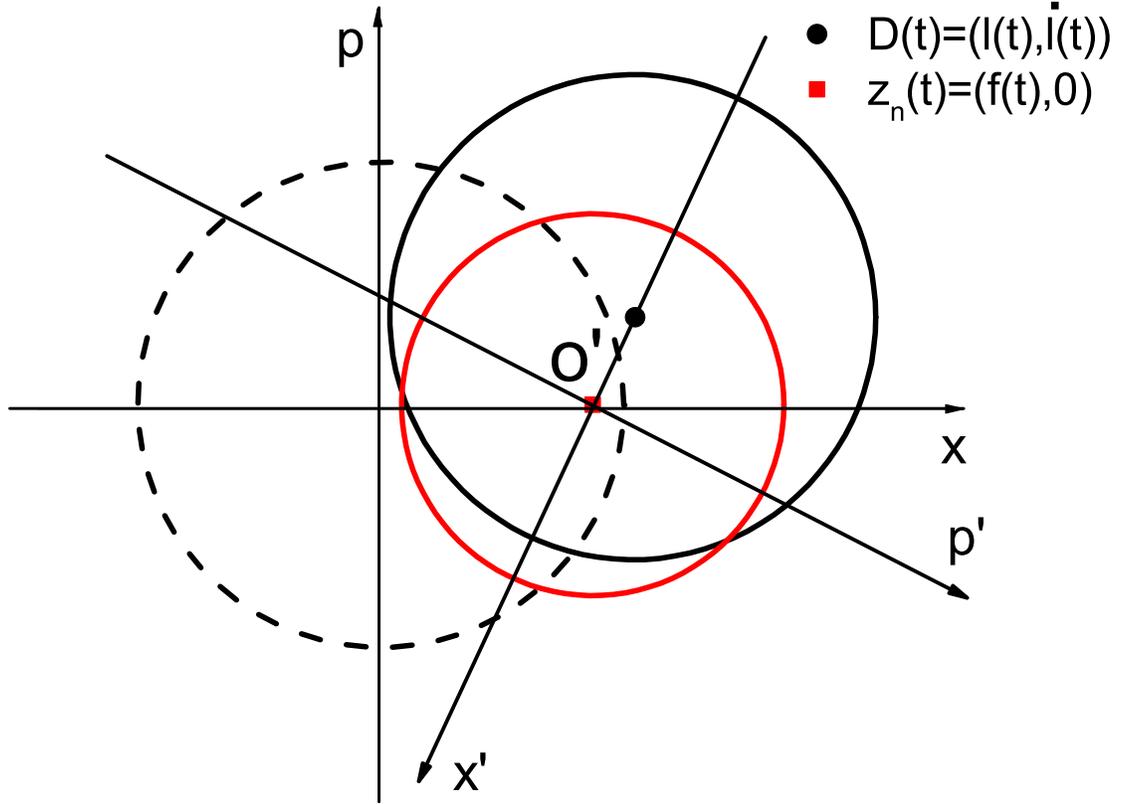}
\caption{{(Color online.) The solid red and black circles represent the two rotation-invariant Weyl symbols, Eqs.~(\ref{Weylsymbolntheigenvectornatural1}) and~(\ref{Weylsymbolntheigenvectornatural2}), in phase space. The quantum transition probability~(\ref{distributioninclusiveworkmicrocanonical}) is equal to their overlap integral. The dashed circle denotes the Weyl symbol of the initial eigenvector $|\varepsilon_m(0)\rangle$. It evolves to the black one at time $t$. } }
\label{fig1}
\end{figure}
Because the integral in~(\ref{distributioninclusiveworkmicrocanonical}) is independent of the coordinate system, we can choose a new coordinate system $(x',p')$ with a new origin $O'$$=$$z_n(t)$. This reminds us that the integral in~(\ref{distributioninclusiveworkmicrocanonical}) is identical to the transition probability between two eigenvectors of two identical simple harmonic oscillators displaced by a distance
\begin{eqnarray}
|D(t)-z_n(t)|&=&\sqrt{\dot{l}(t)^2+[l(t)-f(t)]^2} \nonumber \\
&=&\sqrt{2{\cal W}(t)},
\end{eqnarray}
that is,
\begin{eqnarray}
\label{quantumtransitionprobablityFranckCondon}
P(n,t|m,0)=\left |\int_{-\infty}^{+\infty} dx \phi_n(x)\phi_m\left(x-\sqrt{2{\cal W}(t)}\right) \right|^2,
\end{eqnarray}
where $\phi_{n,m}(x)$$=$$\langle x|\varepsilon_{n,m}\rangle$ is the eigenvector $|\varepsilon_{n,m}\rangle $ of the simple quantum harmonic oscillator~(\ref{standardoscillatorHoperator}) in position representation. The term on the right-hand side of Eq.~(\ref{quantumtransitionprobablityFranckCondon}) is called the Franck-Condon factor in molecular spectra~\cite{Atkins1999} and is equal to~\cite{Hutchisson1930}~\footnote{This result has been reobtained by many authors, e.g., Refs.~\cite{Husimi1953,Dowling1991,Talkner2008}. }
\begin{eqnarray}
\label{weightmicrocanonicalcase}
\frac{1}{n!m!} {\cal W}(t)^{m+n} e^{-{\cal W}(t)}\left\{s! {\cal W}(t)^{-s} L_s^{(|n-m|)}[{\cal W}(t)]\right\}^2,
\end{eqnarray}
where $s$$=$$\min(n,m)$, and $L_s^{(|n-m|)}$ denotes the generalized Laguerre polynomial. The significance of the connection between the quantum transition probability and the Franck-Condon factor is that the latter has an insightful interpretation of ``interference in phase space"~\cite{Wheeler1985,Schleich2001}. Therefore, it provides an alternative route to understand the quantum features of quantum work distributions~\cite{Jarzynski2015}.}

\section{An $\hbar^2$-order numerical method }
\label{section4}
If a quantum system has a potential with powers equal to or greater than three, then we do not have an exact Weyl symbol of the exponential Hamiltonian. Moreover, Eq.~(\ref{vonNeumannequationinphasespace}) is no longer the classical Liouville equation. Hence, we must resort to approximation methods. In our previous work~\cite{Fei2018}, we proved that CF~(\ref{CFworkclosedsysteminphasespace}) can be expanded in a Planck constant series:
\begin{eqnarray}
\label{expandedCF}
\Phi(\eta)= \Phi^{(0)}(\eta) - \sum_{n=1}^\infty \hbar^{2n} \Phi^{(2n)}(\eta),
\end{eqnarray}
where $\Phi^{(0)}(\eta)$ is the CF of the classical work of the classical system. In addition, we also obtained the path-integral representation of the lowest quantum correction, $\Phi^{(2)}(\eta)$. In this section, we attempt to compute the CF approximated to $\hbar^2$ by solving PDEs rather than the dynamic trajectories of the classical system. The latter is required in the path-integral representation of the CF. For simplicity, we focus on the single particle system in a one-dimensional situation~\footnote{{\bf For the systems with degrees of freedom greater than one, the expressions of the $\hbar^2$ terms in Eqs.~(\ref{equationmotioninphasespacetillsecondorder2}) and~(\ref{expandedexponentialHamiltonian}) must be extended to account for the multidimensional case; see the corresponding terms in Eqs. (2.56) and (2.82) of Ref.~\cite{Hillery1984}. Although our approximation method is still applicable, because we face the extra numerical difficulty of solving multidimensional PDEs, we do not consider them in this paper. }}. Therefore, we expand Eq.~(\ref{vonNeumannequationinphasespace}) and the Weyl symbol of the exponential Hamiltonian up to $\hbar^2$~\cite{Wigner1932}:
\begin{eqnarray}
\label{equationmotioninphasespacetillsecondorder2}
{\partial_t P}(z,t)&=&-H(z,t)\Lambda P(z,t) - \hbar^2 \frac{1}{24}{\partial_x^3 }U {\partial_p^3 }P(z,t)+ \cdots, 
\end{eqnarray}
and
\begin{eqnarray}
\label{expandedexponentialHamiltonian}
\left[e^{-i\eta\hat H(t)}\right]_w=e^{-i\eta H(z,t)}\left[1- \hbar^2f(i\eta,z,t)+\cdots\right],
\end{eqnarray}
where
\begin{eqnarray}
\label{fundefinition}
f(i\eta,z,t)=\frac{(i\eta)^2}{8m}\left[\partial_x^2U-\frac{i\eta}{3}(\partial_xU)^2-\frac{i\eta}{3m}p^2\partial_x^2U\right].
\end{eqnarray}
Expanding
\begin{eqnarray}
P(z,t)=P^{(0)}(z,t)-\hbar^2 P^{(2)}(z,t)+\cdots
\end{eqnarray}
as well, substituting it into Eq.~(\ref{equationmotioninphasespacetillsecondorder2}), and collecting all of the terms with the same order of Planck's constant, we obtain
\begin{eqnarray}
\label{PDEs1}
\partial_tP^{(0)}(z,t)&=&-H(z,t)\Lambda P^{(0)}(z,t), \\
\label{PDEs2}
\partial_tP^{(2)}(z,t)&=&-H(z,t)\Lambda P^{(2)}(z,t)+\frac{1}{24}\partial_x^3U(z,t)\partial_p^3P^{(0)}(z,t),
\end{eqnarray}
and their initial conditions are given by
\begin{eqnarray}\label{initialconditionP}
P_0^{(0)}(z)&=&P_{eq}(i\eta+\beta,z,0), \\
P_0^{(2)}(z)&=&P_{eq}(i\eta+\beta,z,0)\delta f(i\eta+\beta,z,0),
\end{eqnarray}
where $P_{eq}(i\eta+\beta,z,0)$ is a complex extension of the classical thermal equilibrium state
\begin{eqnarray}
\label{classicalthermalstate}
P_{eq}(\beta,z,0)=\frac{e^{-\beta H(z,0)}}{\int_{-\infty}^{+\infty} dz e^{-\beta H(z,0)}},
\end{eqnarray}
$\delta f(i\eta+\beta,z,0)$$=$$f(i\eta+\beta,z,0)-\langle f(\beta,0)\rangle_{eq}$, and $\langle f(\beta,0)\rangle_{eq}$ is the average of $f(\beta, z, 0)$ with respect to the distribution function~(\ref{classicalthermalstate}). Solving Eqs.~(\ref{PDEs1}) and~(\ref{PDEs2}), we obtain the approximated or truncated CF~(\ref{expandedCF})
\begin{eqnarray}
\label{expandedCFtill2order}
\Phi(\eta)\approx\Phi^{(0)}(\eta) - \hbar^{2} \Phi^{(2)}(\eta),
\end{eqnarray}
with
\begin{eqnarray}
\label{twoorderCF}
\Phi^{(0)}(\eta)&=&\int_{-\infty}^{+\infty} dz e^{i\eta H(z,t)}P^{(0)}(z,t) , \\
\Phi^{(2)}(\eta)&=&\int_{-\infty}^{+\infty} dz e^{i\eta H(z,t)}\left[ f(-i\eta,z,t)P^{(0)}(z,t) + P^{(2)}(z,t)\right].
\end{eqnarray}
{The first integral on the right-hand side of $\Phi^{(2)}(\eta)$ represents the lowest quantum correction due to the second energy measurement at the end, while the second integral is the sum of the lowest quantum corrections due to the initial quantum state and quantum dynamics. Of course, these interpretations are the same as those regarding the classical path-integral representations. Before we exemplify the approximation method by using concrete quantum models, we want to make two comments on its restrictions. First, the expansion~(\ref{expandedCF}) is essentially an asymptotic series, and it may become nonconvergent for very low temperatures~\cite{Wigner1932,Oppenheim1957,Imre1967} and/or large $\eta$. This point can be seen from the $\hbar$-expansion, Eq.~(\ref{expandedexponentialHamiltonian}) (note the powers of $i\eta$ in Eq.~(\ref{fundefinition})). Second, to obtain the form~(\ref{vonNeumannequationinphasespace}) of the von Neumann equation in phase space, we presuppose that the potential $U(\hat{\bf x},t)$ is spatially smooth enough to have a Taylor expansion~\cite{Wigner1932}. Therefore, for the general potentials with multiple extreme points or potentials whose shapes change dramatically, we do not expect our method to predict satisfactory results. }

The first example is the quantum harmonic oscillator, Eq.~(\ref{Hamiltoniantimemassfrequency}). We set the parameters to be the same as those used by Deffner and Lutz: the mass is a constant, and the frequency varies with time as $\omega_\tau^2$$=$$\omega_0^2+(\omega_t^2-\omega_0^2)\tau/t$ ($0\le \tau\le t$), where $t$ is the duration of the nonequilibrium process. The advantage of this set of parameter values is that Eqs.~(\ref{formalsolution1}) and~(\ref{formalsolution2}) have exact expressions using the Airy functions~\cite{Deffner2008}. Fig.~(\ref{fig1})(a) shows the exact CF~(\ref{CFtimevaryingmassandfrequency}) and the approximated one obtained by numerically solving Eqs.~(\ref{PDEs1}) and~(\ref{PDEs2}). We note that these two PDEs reduce to two independent classical Liouville equations because $\partial_x^3U$$=$$0$ here. We observe that the agreement between these two methods is satisfactory, particularly near $\eta$$=$$0$. {The potential of the harmonic oscillator is quadratic and special. To show the generality of our method, we test it on other driven quantum models with anharmonic potentials. Hence, we choose a family of Hamiltonians with time-dependent even power-law potentials~\cite{Zheng2014},
\begin{eqnarray}
\label{powerlawpotential}
\hat H(t)=\frac{{\hat p}^2}{2m}+\frac{1}{2}\frac{1}{f(t)^2}\left(\frac{\hat{x}}{f(t)}\right)^{2q},
\end{eqnarray}
where $q=1,2,3,4$ and $f(t)$$=$$1+t$. Obviously, $q$ indicates the degree to which a potential deviates from the quadratic potential. The CFs of these systems~(\ref{powerlawpotential}) can be solved by a formally exact method~\cite{Berry1984,Leonard2015}. Fig.~(\ref{fig1})(b) shows a comparison of the data obtained by the exact method and our approximation method. We can see that these data are consistent, even in the case of $q$$=$$4$, where the spatial change of the potential is sharp.

It is important to note that the obvious deviations of these approximated CFs from the exact results at larger $\eta$ do not reduce the significance of our method. The reason for this is that we are truly concerned about the statistics of the quantum work, e.g., the mean and variance. These statistical quantities are determined by taking the $n$th derivatives of the CF with respect to $i\eta$ and $\eta$$=$$0$. In addition, the reader is reminded that the slopes of the imaginary parts of the CFs in Figs.~(\ref{fig1}) (a) and (b) have opposite signs: that is, the mean work done on these two types of systems is positive and negative, respectively. This is expected because the changes in these types of trapping potentials have opposite trends.

\begin{figure}
\includegraphics[angle=0,width=1\linewidth]{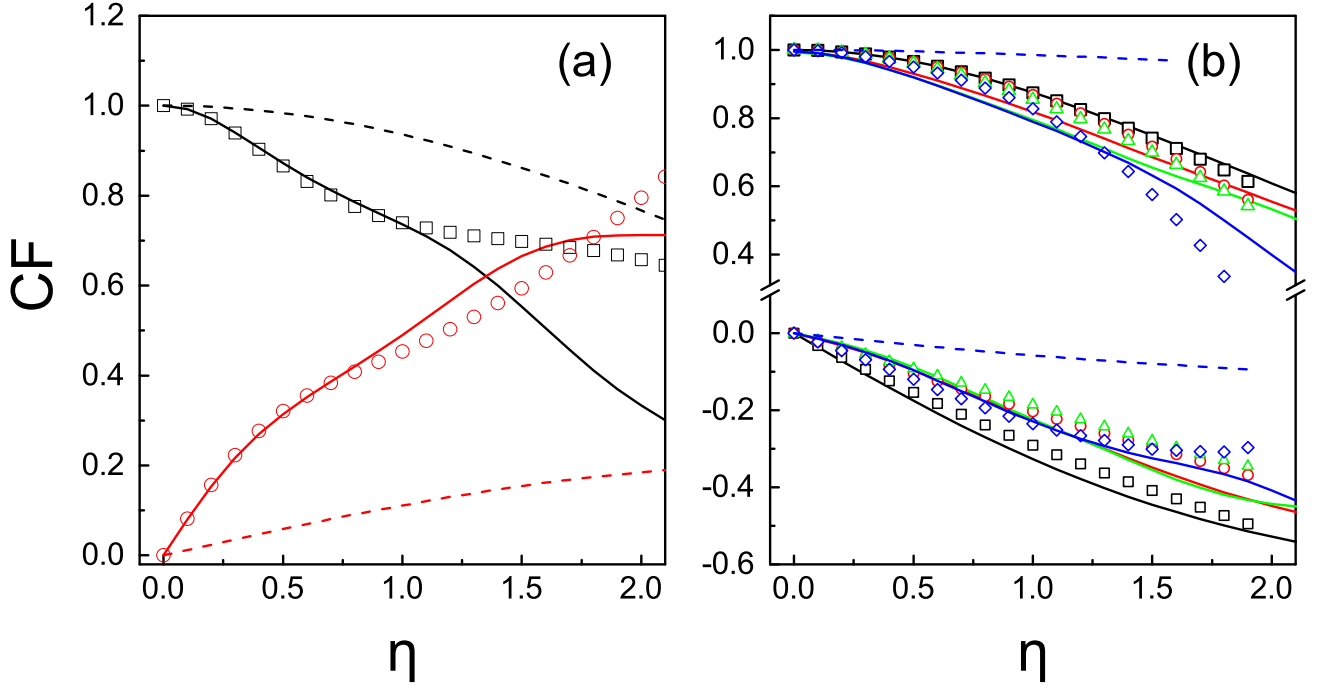}
\caption{{(Color online.) (a). The time-dependent quantum harmonic oscillator: black and red solid lines are the real and imaginary parts, respectively, of the exact CF computed by Eq.~(\ref{CFtimevaryingmassandfrequency}). The black empty squares and red empty circles are the real and imaginary parts of the CFs, respectively, calculated by our method approximated to $\hbar^2$. The parameters used in the model are $\omega_0$$=$$1$, $\omega_1$$=$$2$, $m$$=$$t$$=$$1$, and $\beta$$=$$2$. (b). The driven quantum models with even power potentials: the data starting from the coordinates $(0,1)$ and $(0,0)$ are the real and imaginary parts of the CFs, respectively. The solid curves are the CFs computed by a formally exact method~\cite{Berry1984,Leonard2015}, and the symbols are the CFs obtained by our approximation method. The black, red, green, and blue data correspond to the driven quantum model with $q$$=$$1, 2, 3,$ and $4$, respectively. The parameters used in these models are $m$$=$$t$$=$$1$ and $\beta$$=$$2$. As a comparison, the classical CF, $\Phi^{(0)}(\eta)$, for the driven classical model with $q=4$ is also computed by using the first equation in Eq.~(\ref{twoorderCF}) and is shown by the blue dashed lines. We see in this case that the quantum corrections contribute significantly to the CF. Note that in these two panels, we have set $\hbar$$=$$1$.} }
\label{fig2}
\end{figure}
}

\section{Conclusion}\label{section5}
{In this paper, we obtain several exact CFs of the TEM quantum work of driven systems that consist of quantum harmonic oscillators. Compared to the previous methods that solve the evolution of these quantum systems, our formulas in phase space show unification, and the computations are also simple. The reinvestigation of the quantum harmonic oscillator driven by a time-dependent force reveals a previously unnoticed connection between the quantum work distributions and Franck-Condon factor. For the general quantum systems, obtaining their CFs analytically in phase space seems to be extremely difficult, if not impossible. Therefore, we present a numerical method approximated to $\hbar^2$ order for their CFs and demonstrate its precision using several driven quantum models. Although this approximation method is restricted to situations with smooth driven potentials, small Fourier parameters, and high and moderately low temperatures, we still think that it is valuable. To our knowledge, there are no general strategies to compute the CF of the quantum work for a general quantum system except for direct computation using the definition. Moreover, the mentioned restrictions are usually what quantum thermodynamics researchers are interested in. There are two possible extensions of this study. One is to quantitatively understand the quantum features of quantum work distributions from the perspective of interference in phase space. The other is determining how to take quantum Bose or Fermi statistics into account when we study the quantum work statistics of many particle systems in phase space. We hope to report these results in the near future.}

\begin{acknowledgments}
This work was supported by the National Science Foundation of China under grant nos. 11174025 and 11575016. We also appreciate the support of the CAS Interdisciplinary Innovation Team, No. 2060299.
\end{acknowledgments}

\appendix
\section{Some useful formulas in Sec.~(\ref{section3})}
Substituting the Weyl symbol of the exponential Eq.~(\ref{WignerOmega}) into the CF~(\ref{CFworkclosedsysteminphasespaceoscillator}), we have
\begin{eqnarray}
\label{CFdetaildrivingoscillator}
\Phi(\eta)&=&\frac{1}{\pi\hbar} {\rm sech}\left(\frac{(i\eta+\beta)\hbar\omega}{2}\right) {\rm sech}\left(\frac{i\hbar\eta\omega}{2}\right)
\sinh\left( \frac{\hbar\omega\beta}{2}\right)\exp\left[\frac{-i\eta F^2(t)}{2m\omega^2}\right]\nonumber \\
&&\int_{-\infty}^{+\infty} dz_0\exp\left[ \frac{2}{\hbar\omega}\tanh\left(\frac{i\eta\hbar\omega}{2}\right)H_F'(z_t,t) - \frac{2}{\hbar\omega}\tanh\left(\frac{(\beta+i\eta)\hbar\omega}{2}\right)H_F'(z_0,0) \right].
\end{eqnarray}
The dynamic solution $z_t^T$$=$$(x_t,p_t)$ of the classical harmonic oscillator with the Hamiltonian
\begin{eqnarray}
H_F(z,t)=\frac{p^2}{2m}+\frac{m\omega^2x^2}{2} -F(t)x
\end{eqnarray}
is simply
\begin{eqnarray}
\label{dynamicsolutiondrivingoscillator1}
x_t&=& \cos(\omega t) x_0+\frac{\sin(\omega t)}{m\omega} p_0 +l(t),\\
\label{dynamicsolutiondrivingoscillator2}
p_t&=&-m \omega \sin(\omega t)x_0+\cos(\omega t) p_0 + m\dot{l}(t),
\end{eqnarray}
where $z_0^T$$=$$(x_0,p_0)$ is the initial phase point and the function $l(t)$ is
\begin{eqnarray}
\label{definitionoflfunction}
\frac{1}{m\omega}\int_0^t F(s)\sin(\omega(t-s))ds.
\end{eqnarray}
{Eqs.~(\ref{dynamicsolutiondrivingoscillator1}) and~(\ref{dynamicsolutiondrivingoscillator2}) can be rewritten into a compact form by using vector and matrix notations:
\begin{eqnarray}
\label{vectorformofdynamicalsolution}
{ z}_t=R(t){ z}_0 + D(t).
\end{eqnarray}
${D}^T(t)$$=$$(l,m\dot{l})$ denotes the phase point of the classical oscillator moving under the time-dependent force $F(t)$, which is initially at rest. }Substituting the solution into $H_F'(z_t,t)$ in Eq.~(\ref{CFdetaildrivingoscillator}), expanding all functions $H_F'$ in terms of $x_0$ and $p_0$ and rearranging, we find that the integral is Gaussian with respect to the two variables $x_0$ and $p_0$. Then, we arrive at Eq.~(\ref{CFharmonicoscillatortimedependentforce}). {If we further expand the hyperbolic functions therein into exponential functions, we obtain an expression found by Talkner et al.~\cite{Talkner2008}:
\begin{eqnarray}
\label{Talknerequation}
\exp\left[-\frac{i\eta F^2(t) }{2m\omega^2}+\frac{{\cal W}(t)}{\hbar\omega}{\left(e^{i\eta\hbar\omega} -1\right)} -4\frac{{\cal W}(t)}{\hbar\omega} \frac{\sin^2(\hbar\omega\eta/2)}{\left(e^{\beta\hbar\omega}-1\right)} \right].
\end{eqnarray}
The reason that we call ${\cal W}(t)$ the work done on Hamiltonian system~$H_F'(z,t)$ initially at rest is that
\begin{eqnarray}
H'_F(z_t,t)-H'_F(z_0,0)=a(t)x_0+b(t)p_0+{\cal W}(t),
\end{eqnarray}
where $a(t)$ and $b(t)$ are some functions of time $t$, and
\begin{eqnarray}
\label{classicalwork2}
{\cal W}(t) = \frac{1}{2m\omega^2}\left[\left(m \omega^2 l(t)-F(t)\right)^2 + (m\omega\dot{l}(t))^2 \right].
\end{eqnarray}
The equivalence between the above equation and Eq.~(\ref{classicalwork}) can be verified directly. Considering that these results are elementary and have been presented in the previous literature~\cite{Husimi1953,Campisi2008,Fei2018}, we do not provide further details.
}

The above procedure is also applicable to the quantum harmonic oscillator with time-dependent mass and frequency, Eq.~(\ref{Hamiltoniantimemassfrequency}). Here, we must apply the Weyl symbol Eq.~(\ref{WeylexponentialHamiltonianoscillator})~\footnote{Now the mass and frequency are time dependent.} and the formal dynamic solutions~(\ref{formalsolution1}) and~(\ref{formalsolution2}). To obtain Eq.~(\ref{CFtimevaryingmassandfrequency}), we have expanded all possible hyperbolic functions into exponential functions and used the Wronskian determinant
\begin{eqnarray}
\label{Wronskiandet}
X(t)R(t)-Y(t)Q(t)=1.
\end{eqnarray}
We can verify Eq.~(\ref{Wronskiandet}) by noting that $(X(t),Q(t))$ and $(Y(t),R(t))$ are the dynamic solutions of the classical system satisfying the specific initial phase points $(1,0)$ and $(0,1)$, respectively.

{Substituting Eqs.~(\ref{WignerOmega}) and~(\ref{Weylsymbolntheigenvector}) into Eq.~(\ref{CFworkclosedsysteminphasespaceoscillatormicrocanoical}), we obtain
\begin{eqnarray}
\label{CFworkclosedsysteminphasespaceoscillatormicrocanoical2}
\Phi_m(\eta)&=&\frac{(-1)^n}{\pi\hbar}\exp\left[-\frac{i\eta F(t)^2}{2m\omega^2}\right]
{\rm sech} \left( \frac{i\eta\hbar\omega}{2}\right)
\exp \left[ -i\eta \left( n +\frac{1}{2} \right)\hbar\omega\right] \nonumber \\
&&\int_{-\infty}^{+\infty} d{z}_0 \exp\left[\frac{2}{\hbar\omega}{\rm tanh}\left(\frac{i\eta\hbar\omega }{2}\right)H_F'(z_t,t)\right]\exp\left[ -\frac{2H_F'(z_0,0)}{\hbar\omega}\right]L_n\left[\frac{4H_F'(z_0,0)}{\hbar\omega} \right] .
\end{eqnarray}
Using the generating function of the Laguerre polynomial,
\begin{eqnarray}
\label{generatingfunction}
\sum_{n=0}^\infty \alpha^n L_n(x)=\frac{1}{1-\alpha}e^{ -\alpha x/( 1-\alpha) },
\end{eqnarray}
we can construct a generating function about $\Phi_m(\eta)$ and obtain
\begin{eqnarray}
\label{generatingfunctionCF}
\sum_{n=0}^\infty \alpha^n\Phi_n(\eta) =\exp\left[-\frac{i\eta F(t)^2}{2m\omega^2}\right] \exp\left[\frac{{\cal W}(t)}{\hbar\omega}\left(e^{i\eta\hbar\omega}-1\right)\right] \frac{1}{1-\alpha}\exp\left[- \frac{\alpha}{1-\alpha} \frac{4{\cal W}(t)}{\hbar\omega} \sin^2\left(\frac{\eta\hbar\omega}{2} \right) \right].
\end{eqnarray}
This derivation is highly analogous to that of Eq.~(\ref{CFharmonicoscillatortimedependentforce}). When we compare Eqs.~(\ref{generatingfunction}) with~(\ref{generatingfunctionCF}), we immediately obtain Eq.~(\ref{CFmicrocanonicalinitialstate}). According to the definition of CF, Eq.~(\ref{CFworkclosedsystem}), there is a simple relation between the CFs of quantum systems with thermal initial state and with microcanonical initial state:
\begin{eqnarray}
\Phi(\eta)=\sum_m P_m(0)\Phi_m(\eta),
\end{eqnarray}
where $P_m(0)$$=$$\exp[-\beta\varepsilon_m(0)]/{\cal Z}_0$. Hence, the latter is more fundamental than the former. }

\section{Weyl symbol of the exponential Hamiltonian~(\ref{twooscillatorhamiltonian}) }
We did not find reports in the literature on the derivation of the Weyl symbol of the exponential Hamiltonian of coupled harmonic oscillators. Let the operator $\hat\Xi(\eta)$$\equiv$$\exp[i\eta\hat H(t)]$ and its Weyl symbol be $\Xi({\bf z},\eta)$. The operator satisfies a Schr{\"o}dinger-like equation~\footnote{If $i\eta\rightarrow -\beta$, then the equation is named the Bloch equation~\cite{Imre1967,Oppenheim1957}}:
\begin{eqnarray}
\label{Schrodingerlikeeqoperator}
i\partial_\eta \hat\Xi(\eta) =- \hat H (t)\hat\Xi(\eta),
\end{eqnarray}
and the initial condition is $\hat\Xi(\eta=0)$$=$$\hat I$, where $\hat I$ is the identity operator. In phase space, the above equation is written as~\cite{Oppenheim1957,Imre1967,Hillery1984}
\begin{eqnarray}
\label{Schrodingerlikeeq}
i\partial_\eta \Xi({\bf z},\eta) =- H({\bf z},t) \cos\left(\frac{i\hbar}{2}\Lambda \right)\Xi({\bf z},\eta).
\end{eqnarray}
It is well-known that under the canonical transformation, the Poisson bracket is invariant, that is, $\Lambda$$=$$\sum_{i=1}^{2} \Lambda_i$, where
\begin{eqnarray}
\Lambda_i= \overleftarrow\partial_{P_i}\overrightarrow\partial_{Q_i}-\overleftarrow\partial_{Q_i}\overrightarrow\partial_{P_i}.
\end{eqnarray}
Hence, after the transformation, Eq.~(\ref{Schrodingerlikeeq}) becomes
\begin{eqnarray}
\label{Weylsymboltwoparticles}
i\partial_\eta \Xi({\bf Z},\eta)&=&- \left[\sum _{i=1}^2H_i({\bf Z}_i , t) \right]\cos\left(\frac{i\hbar}{2}\Lambda \right)\Xi({\bf Z},\eta)\nonumber \\
&=&- \left[\sum _{i=1}^2H_i({\bf Z}_i , t)\cos\left(\frac{i\hbar}{2}\Lambda_i \right) \right]\Xi({\bf Z},\eta).
\end{eqnarray}
The reason for the second equation is that unless the phase coordinates are the same, the action of Poisson bracket $\Lambda_i$ on Hamiltonian $H_j$ ($j$$\neq$$i$) is always zero. Then, we can readily verify that the solution of Eq.~(\ref{Weylsymboltwoparticles}) is
\begin{eqnarray}
\Xi({\bf z},\eta)=\prod_{i=1}^2 \Xi({\bf Z}_i, \eta),
\end{eqnarray}
where $\Xi({\bf Z}_i, \eta)$ is Eq.~(\ref{WignerOmega}) for the specified Hamiltonian operator~(\ref{Hindependentoscillators}).
In addition, the partition function can also be calculated by the canonical transformation
\begin{eqnarray}
\label{partitionnoscillators}
{\cal Z}_0&=&\frac{1}{(2\pi\hbar)^2}\int_{-\infty}^{+\infty} d{\bf z} \Xi({\bf z},i\beta) \nonumber \\
&=&\prod_{i=1}^2\frac{1}{2\pi\hbar}\int_{-\infty}^{+\infty} d{\bf Z}_i \Xi({\bf Z}_i,i\beta)\nonumber \\
&=&\prod_{i=1}^2 {\cal Z}_i,
\end{eqnarray}
where $Z_i$ is the partition function of the quantum Hamiltonian operator~(\ref{Hindependentoscillators}).
In the second equation, we have used the fact that under the canonical transformation, the Jacobian determinant $|d{\bf z}/d{\bf Z}|$ equals $1$. Clearly, Eqs.~(\ref{Schrodingerlikeeqoperator})-~(\ref{partitionnoscillators}) are still valid for the multi-harmonic oscillator case described by Eq.~(\ref{multiharmonicoscillators}). The only necessary change is the replacement of $2$ in these equations by the degrees of freedom of these oscillators.


\begin{thebibliography}{45}%
\makeatletter
\providecommand \@ifxundefined [1]{%
 \@ifx{#1\undefined}
}%
\providecommand \@ifnum [1]{%
 \ifnum #1\expandafter \@firstoftwo
 \else \expandafter \@secondoftwo
 \fi
}%
\providecommand \@ifx [1]{%
 \ifx #1\expandafter \@firstoftwo
 \else \expandafter \@secondoftwo
 \fi
}%
\providecommand \natexlab [1]{#1}%
\providecommand \enquote  [1]{``#1''}%
\providecommand \bibnamefont  [1]{#1}%
\providecommand \bibfnamefont [1]{#1}%
\providecommand \citenamefont [1]{#1}%
\providecommand \href@noop [0]{\@secondoftwo}%
\providecommand \href [0]{\begingroup \@sanitize@url \@href}%
\providecommand \@href[1]{\@@startlink{#1}\@@href}%
\providecommand \@@href[1]{\endgroup#1\@@endlink}%
\providecommand \@sanitize@url [0]{\catcode `\\12\catcode `\$12\catcode
  `\&12\catcode `\#12\catcode `\^12\catcode `\_12\catcode `\%12\relax}%
\providecommand \@@startlink[1]{}%
\providecommand \@@endlink[0]{}%
\providecommand \url  [0]{\begingroup\@sanitize@url \@url }%
\providecommand \@url [1]{\endgroup\@href {#1}{\urlprefix }}%
\providecommand \urlprefix  [0]{URL }%
\providecommand \Eprint [0]{\href }%
\providecommand \doibase [0]{http://dx.doi.org/}%
\providecommand \selectlanguage [0]{\@gobble}%
\providecommand \bibinfo  [0]{\@secondoftwo}%
\providecommand \bibfield  [0]{\@secondoftwo}%
\providecommand \translation [1]{[#1]}%
\providecommand \BibitemOpen [0]{}%
\providecommand \bibitemStop [0]{}%
\providecommand \bibitemNoStop [0]{.\EOS\space}%
\providecommand \EOS [0]{\spacefactor3000\relax}%
\providecommand \BibitemShut  [1]{\csname bibitem#1\endcsname}%
\let\auto@bib@innerbib\@empty
\bibitem [{\citenamefont {Jarzynski}(1997)}]{Jarzynski1997}%
  \BibitemOpen
  \bibfield  {author} {\bibinfo {author} {\bibfnamefont {C.}~\bibnamefont
  {Jarzynski}},\ }\href {\doibase 10.1103/PhysRevLett.78.2690} {\bibfield
  {journal} {\bibinfo  {journal} {Phys. Rev. Lett.}\ }\textbf {\bibinfo
  {volume} {78}},\ \bibinfo {pages} {2690} (\bibinfo {year}
  {1997})}\BibitemShut {NoStop}%
\bibitem [{\citenamefont {Crooks}(1999)}]{Crooks1999}%
  \BibitemOpen
  \bibfield  {author} {\bibinfo {author} {\bibfnamefont {G.~E.}\ \bibnamefont
  {Crooks}},\ }\href {http://pre.aps.org/abstract/PRE/v60/i3/p2721_1}
  {\bibfield  {journal} {\bibinfo  {journal} {Phys. Rev.E}\ }\textbf {\bibinfo
  {volume} {60}},\ \bibinfo {pages} {2721} (\bibinfo {year}
  {1999})}\BibitemShut {NoStop}%
\bibitem [{\citenamefont {Kurchan}(2000)}]{Kurchan2000}%
  \BibitemOpen
  \bibfield  {author} {\bibinfo {author} {\bibfnamefont {J.}~\bibnamefont
  {Kurchan}},\ }\href {http://arxiv.org/abs/cond-mat/0007360} {\bibfield
  {journal} {\bibinfo  {journal} {arXiv preprint cond-mat/0007360}\ } (\bibinfo
  {year} {2000})}\BibitemShut {NoStop}%
\bibitem [{\citenamefont {Tasaki}(2000)}]{Tasaki2000}%
  \BibitemOpen
  \bibfield  {author} {\bibinfo {author} {\bibfnamefont {H.}~\bibnamefont
  {Tasaki}},\ }\href {http://arxiv.org/abs/cond-mat/0009244} {\bibfield
  {journal} {\bibinfo  {journal} {arXiv preprint cond-mat/0009244}\ } (\bibinfo
  {year} {2000})}\BibitemShut {NoStop}%
\bibitem [{\citenamefont {Talkner}\ \emph {et~al.}(2007)\citenamefont
  {Talkner}, \citenamefont {Lutz},\ and\ \citenamefont
  {H\"anggi}}]{Talkner2007}%
  \BibitemOpen
  \bibfield  {author} {\bibinfo {author} {\bibfnamefont {P.}~\bibnamefont
  {Talkner}}, \bibinfo {author} {\bibfnamefont {E.}~\bibnamefont {Lutz}}, \
  and\ \bibinfo {author} {\bibfnamefont {P.}~\bibnamefont {H\"anggi}},\ }\href
  {\doibase 10.1103/PhysRevE.75.050102} {\bibfield  {journal} {\bibinfo
  {journal} {Phys. Rev. E}\ }\textbf {\bibinfo {volume} {75}},\ \bibinfo
  {pages} {050102(R)} (\bibinfo {year} {2007})}\BibitemShut {NoStop}%
\bibitem [{\citenamefont {Campisi}\ \emph {et~al.}(2011)\citenamefont
  {Campisi}, \citenamefont {H{\"a}nggi},\ and\ \citenamefont
  {Talkner}}]{Campisi2011}%
  \BibitemOpen
  \bibfield  {author} {\bibinfo {author} {\bibfnamefont {M.}~\bibnamefont
  {Campisi}}, \bibinfo {author} {\bibfnamefont {P.}~\bibnamefont {H{\"a}nggi}},
  \ and\ \bibinfo {author} {\bibfnamefont {P.}~\bibnamefont {Talkner}},\ }\href
  {http://rmp.aps.org/abstract/RMP/v83/i3/p771_1} {\bibfield  {journal}
  {\bibinfo  {journal} {Rev. Mod. Phys.}\ }\textbf {\bibinfo {volume} {83}},\
  \bibinfo {pages} {771} (\bibinfo {year} {2011})}\BibitemShut {NoStop}%
\bibitem [{\citenamefont {Esposito}\ \emph {et~al.}(2009)\citenamefont
  {Esposito}, \citenamefont {Harbola},\ and\ \citenamefont
  {Mukamel}}]{Esposito2009}%
  \BibitemOpen
  \bibfield  {author} {\bibinfo {author} {\bibfnamefont {M.}~\bibnamefont
  {Esposito}}, \bibinfo {author} {\bibfnamefont {U.}~\bibnamefont {Harbola}}, \
  and\ \bibinfo {author} {\bibfnamefont {S.}~\bibnamefont {Mukamel}},\ }\href
  {http://rmp.aps.org/abstract/RMP/v81/i4/p1665_1} {\bibfield  {journal}
  {\bibinfo  {journal} {Rev. Mod. Phys.}\ }\textbf {\bibinfo {volume} {81}},\
  \bibinfo {pages} {1665} (\bibinfo {year} {2009})}\BibitemShut {NoStop}%
\bibitem [{\citenamefont {Liu}(2018)}]{Liu2018}%
  \BibitemOpen
  \bibfield  {author} {\bibinfo {author} {\bibfnamefont {F.}~\bibnamefont
  {Liu}},\ }\href {\doibase 10.13725/j.cnki.pip.2018.01.001} {\bibfield
  {journal} {\bibinfo  {journal} {Prog. Phys.}\ }\textbf {\bibinfo {volume}
  {38}},\ \bibinfo {pages} {1} (\bibinfo {year} {2018})}\BibitemShut {NoStop}%
\bibitem [{\citenamefont {Batalh{\~{a}}o}\ \emph {et~al.}(2014)\citenamefont
  {Batalh{\~{a}}o}, \citenamefont {Souza}, \citenamefont {Mazzola},
  \citenamefont {Auccaise}, \citenamefont {Sarthour}, \citenamefont {Oliveira},
  \citenamefont {Goold}, \citenamefont {{De Chiara}}, \citenamefont
  {Paternostro},\ and\ \citenamefont {Serra}}]{Batalhao2014}%
  \BibitemOpen
  \bibfield  {author} {\bibinfo {author} {\bibfnamefont {T.~B.}\ \bibnamefont
  {Batalh{\~{a}}o}}, \bibinfo {author} {\bibfnamefont {A.~M.}\ \bibnamefont
  {Souza}}, \bibinfo {author} {\bibfnamefont {L.}~\bibnamefont {Mazzola}},
  \bibinfo {author} {\bibfnamefont {R.}~\bibnamefont {Auccaise}}, \bibinfo
  {author} {\bibfnamefont {R.~S.}\ \bibnamefont {Sarthour}}, \bibinfo {author}
  {\bibfnamefont {I.~S.}\ \bibnamefont {Oliveira}}, \bibinfo {author}
  {\bibfnamefont {J.}~\bibnamefont {Goold}}, \bibinfo {author} {\bibfnamefont
  {G.}~\bibnamefont {{De Chiara}}}, \bibinfo {author} {\bibfnamefont
  {M.}~\bibnamefont {Paternostro}}, \ and\ \bibinfo {author} {\bibfnamefont
  {R.~M.}\ \bibnamefont {Serra}},\ }\href@noop {} {\bibfield  {journal}
  {\bibinfo  {journal} {Phys. Rev. Lett.}\ }\textbf {\bibinfo {volume} {113}},\
  \bibinfo {pages} {140601} (\bibinfo {year} {2014})}\BibitemShut {NoStop}%
\bibitem [{\citenamefont {An}\ \emph {et~al.}(2015)\citenamefont {An},
  \citenamefont {Zhang}, \citenamefont {Um}, \citenamefont {Lv}, \citenamefont
  {Lu}, \citenamefont {Zhang}, \citenamefont {Yi}, \citenamefont {Quan},\ and\
  \citenamefont {Kim}}]{ShuomingAn2015}%
  \BibitemOpen
  \bibfield  {author} {\bibinfo {author} {\bibfnamefont {S.-M.}\ \bibnamefont
  {An}}, \bibinfo {author} {\bibfnamefont {J.-N.}\ \bibnamefont {Zhang}},
  \bibinfo {author} {\bibfnamefont {M.}~\bibnamefont {Um}}, \bibinfo {author}
  {\bibfnamefont {D.-S.}\ \bibnamefont {Lv}}, \bibinfo {author} {\bibfnamefont
  {Y.}~\bibnamefont {Lu}}, \bibinfo {author} {\bibfnamefont {J.-H.}\
  \bibnamefont {Zhang}}, \bibinfo {author} {\bibfnamefont {Z.-Q.}\ \bibnamefont
  {Yi}}, \bibinfo {author} {\bibfnamefont {H.-T.}\ \bibnamefont {Quan}}, \ and\
  \bibinfo {author} {\bibfnamefont {K.}~\bibnamefont {Kim}},\ }\href@noop {}
  {\bibfield  {journal} {\bibinfo  {journal} {Nat. Phys.}\ }\textbf {\bibinfo
  {volume} {11}},\ \bibinfo {pages} {193} (\bibinfo {year} {2015})}\BibitemShut
  {NoStop}%
\bibitem [{\citenamefont {Allahverdyan}\ and\ \citenamefont
  {Nieuwenhuizen}(2005)}]{Allahverdyan2005}%
  \BibitemOpen
  \bibfield  {author} {\bibinfo {author} {\bibfnamefont {A.~E.}\ \bibnamefont
  {Allahverdyan}}\ and\ \bibinfo {author} {\bibfnamefont {T.~M.}\ \bibnamefont
  {Nieuwenhuizen}},\ }\href
  {http://journals.aps.org/pre/abstract/10.1103/PhysRevE.71.066102} {\bibfield
  {journal} {\bibinfo  {journal} {Phys. Rev. E}\ }\textbf {\bibinfo {volume}
  {71}},\ \bibinfo {pages} {066102} (\bibinfo {year} {2005})}\BibitemShut
  {NoStop}%
\bibitem [{\citenamefont {B\"{a}umer}\ \emph {et~al.}(2018)\citenamefont
  {B\"{a}umer}, \citenamefont {Lostaglio}, \citenamefont {Perarnau-Llobet},\
  and\ \citenamefont {Sampaio}}]{Bumer2018}%
  \BibitemOpen
  \bibfield  {author} {\bibinfo {author} {\bibfnamefont {E.}~\bibnamefont
  {B\"{a}umer}}, \bibinfo {author} {\bibfnamefont {M.}~\bibnamefont
  {Lostaglio}}, \bibinfo {author} {\bibfnamefont {M.}~\bibnamefont
  {Perarnau-Llobet}}, \ and\ \bibinfo {author} {\bibfnamefont {R.}~\bibnamefont
  {Sampaio}},\ }in\ \href {\doibase 10.1007/978-3-319-99046-0_11} {\emph
  {\bibinfo {booktitle} {Fundamental Theories of Physics}}}\ (\bibinfo
  {publisher} {Springer International Publishing},\ \bibinfo {year} {2018})\
  pp.\ \bibinfo {pages} {275--300}\BibitemShut {NoStop}%
\bibitem [{\citenamefont {Jarzynski}\ \emph {et~al.}(2015)\citenamefont
  {Jarzynski}, \citenamefont {Quan},\ and\ \citenamefont
  {Rahav}}]{Jarzynski2015}%
  \BibitemOpen
  \bibfield  {author} {\bibinfo {author} {\bibfnamefont {C.}~\bibnamefont
  {Jarzynski}}, \bibinfo {author} {\bibfnamefont {H.T.}~\bibnamefont {Quan}}, \
  and\ \bibinfo {author} {\bibfnamefont {S.}~\bibnamefont {Rahav}},\
  }\href@noop {} {\bibfield  {journal} {\bibinfo  {journal} {Phys. Rev. X}\
  }\textbf {\bibinfo {volume} {5}},\ \bibinfo {pages} {031038} (\bibinfo {year}
  {2015})}\BibitemShut {NoStop}%
\bibitem [{\citenamefont {Zhu}\ \emph {et~al.}(2016)\citenamefont {Zhu},
  \citenamefont {Gong}, \citenamefont {Wu},\ and\ \citenamefont
  {Quan}}]{Zhu2016}%
  \BibitemOpen
  \bibfield  {author} {\bibinfo {author} {\bibfnamefont {L.}~\bibnamefont
  {Zhu}}, \bibinfo {author} {\bibfnamefont {Z.}~\bibnamefont {Gong}}, \bibinfo
  {author} {\bibfnamefont {B.}~\bibnamefont {Wu}}, \ and\ \bibinfo {author}
  {\bibfnamefont {H.~T.}\ \bibnamefont {Quan}},\ }\href {\doibase
  10.1103/PhysRevE.93.062108} {\bibfield  {journal} {\bibinfo  {journal} {Phys.
  Rev. E}\ }\textbf {\bibinfo {volume} {93}},\ \bibinfo {pages} {062108}
  (\bibinfo {year} {2016})}\BibitemShut {NoStop}%
\bibitem [{\citenamefont {Wang}\ and\ \citenamefont {Quan}(2017)}]{Wang2017a}%
  \BibitemOpen
  \bibfield  {author} {\bibinfo {author} {\bibfnamefont {Q.}~\bibnamefont
  {Wang}}\ and\ \bibinfo {author} {\bibfnamefont {H.~T.}\ \bibnamefont
  {Quan}},\ }\href {\doibase 10.1103/PhysRevE.95.032113} {\bibfield  {journal}
  {\bibinfo  {journal} {Phys. Rev. E}\ }\textbf {\bibinfo {volume} {95}},\
  \bibinfo {pages} {032113} (\bibinfo {year} {2017})}\BibitemShut {NoStop}%
\bibitem [{\citenamefont {Garc\'{\i}a-Mata}\ \emph {et~al.}(2017)\citenamefont
  {Garc\'{\i}a-Mata}, \citenamefont {Roncaglia},\ and\ \citenamefont
  {Wisniacki}}]{Garcia-Mata2017}%
  \BibitemOpen
  \bibfield  {author} {\bibinfo {author} {\bibfnamefont {I.}~\bibnamefont
  {Garc\'{\i}a-Mata}}, \bibinfo {author} {\bibfnamefont {A.~J.}\ \bibnamefont
  {Roncaglia}}, \ and\ \bibinfo {author} {\bibfnamefont {D.~A.}\ \bibnamefont
  {Wisniacki}},\ }\href {\doibase 10.1103/PhysRevE.95.050102} {\bibfield
  {journal} {\bibinfo  {journal} {Phys. Rev. E}\ }\textbf {\bibinfo {volume}
  {95}},\ \bibinfo {pages} {050102(R)} (\bibinfo {year} {2017})}\BibitemShut
  {NoStop}%
\bibitem [{\citenamefont {Fei}\ \emph {et~al.}(2018)\citenamefont {Fei},
  \citenamefont {Quan},\ and\ \citenamefont {Liu}}]{Fei2018}%
  \BibitemOpen
  \bibfield  {author} {\bibinfo {author} {\bibfnamefont {Z.}~\bibnamefont
  {Fei}}, \bibinfo {author} {\bibfnamefont {H.~T.}\ \bibnamefont {Quan}}, \
  and\ \bibinfo {author} {\bibfnamefont {F.}~\bibnamefont {Liu}},\ }\href
  {\doibase 10.1103/PhysRevE.98.012132} {\bibfield  {journal} {\bibinfo
  {journal} {Phys. Rev. E}\ }\textbf {\bibinfo {volume} {98}},\ \bibinfo
  {pages} {012132} (\bibinfo {year} {2018})}\BibitemShut {NoStop}%
\bibitem [{\citenamefont {Wigner}(1932)}]{Wigner1932}%
  \BibitemOpen
  \bibfield  {author} {\bibinfo {author} {\bibfnamefont {E.}~\bibnamefont
  {Wigner}},\ }\href {\doibase 10.1103/PhysRev.40.749} {\bibfield  {journal}
  {\bibinfo  {journal} {Phys. Rev.}\ }\textbf {\bibinfo {volume} {40}},\
  \bibinfo {pages} {749} (\bibinfo {year} {1932})}\BibitemShut {NoStop}%
\bibitem [{\citenamefont {Landau}\ and\ \citenamefont
  {Lifshits}(1997)}]{LandauStatistics}%
  \BibitemOpen
  \bibfield  {author} {\bibinfo {author} {\bibfnamefont {L.~D.}\ \bibnamefont
  {Landau}}\ and\ \bibinfo {author} {\bibfnamefont {E.~M.}\ \bibnamefont
  {Lifshits}},\ }\href@noop {} {\emph {\bibinfo {title} {Statistical Physics
  (Part I)}}},\ \bibinfo {edition} {3rd}\ ed.\ (\bibinfo  {publisher} {Oxford:
  Butterworth-Heinemann},\ \bibinfo {year} {1997})\BibitemShut {NoStop}%
\bibitem [{\citenamefont {Husimi}(1953)}]{Husimi1953}%
  \BibitemOpen
  \bibfield  {author} {\bibinfo {author} {\bibfnamefont {K.}~\bibnamefont
  {Husimi}},\ }\href@noop {} {\bibfield  {journal} {\bibinfo  {journal} {Prog.
  Theor. Phys.}\ }\textbf {\bibinfo {volume} {9}},\ \bibinfo {pages} {238}
  (\bibinfo {year} {1953})}\BibitemShut {NoStop}%
\bibitem [{\citenamefont {Pedrosa}(1997)}]{Pedrosa1997}%
  \BibitemOpen
  \bibfield  {author} {\bibinfo {author} {\bibfnamefont {I.~A.}\ \bibnamefont
  {Pedrosa}},\ }\href {\doibase 10.1103/PhysRevA.55.3219} {\bibfield  {journal}
  {\bibinfo  {journal} {Phys. Rev. A}\ }\textbf {\bibinfo {volume} {55}},\
  \bibinfo {pages} {3219} (\bibinfo {year} {1997})}\BibitemShut {NoStop}%
\bibitem [{\citenamefont {Hillery}\ \emph {et~al.}(1984)\citenamefont
  {Hillery}, \citenamefont {O'Connell}, \citenamefont {Scully},\ and\
  \citenamefont {Wigner}}]{Hillery1984}%
  \BibitemOpen
  \bibfield  {author} {\bibinfo {author} {\bibfnamefont {M.}~\bibnamefont
  {Hillery}}, \bibinfo {author} {\bibfnamefont {R.}~\bibnamefont {O'Connell}},
  \bibinfo {author} {\bibfnamefont {M.}~\bibnamefont {Scully}}, \ and\ \bibinfo
  {author} {\bibfnamefont {E.}~\bibnamefont {Wigner}},\ }\href {\doibase
  https://doi.org/10.1016/0370-1573(84)90160-1} {\bibfield  {journal} {\bibinfo
   {journal} {Phys. Rep.}\ }\textbf {\bibinfo {volume} {106}},\ \bibinfo
  {pages} {121 } (\bibinfo {year} {1984})}\BibitemShut {NoStop}%
\bibitem [{\citenamefont {Polkovnikov}(2010)}]{Polkovnikov2010}%
  \BibitemOpen
  \bibfield  {author} {\bibinfo {author} {\bibfnamefont {A.}~\bibnamefont
  {Polkovnikov}},\ }\href {\doibase https://doi.org/10.1016/j.aop.2010.02.006}
  {\bibfield  {journal} {\bibinfo  {journal} {Ann. Phys.}\ }\textbf {\bibinfo
  {volume} {325}},\ \bibinfo {pages} {1790 } (\bibinfo {year}
  {2010})}\BibitemShut {NoStop}%
\bibitem [{\citenamefont {Schleich}(2001)}]{Schleich2001}%
  \BibitemOpen
  \bibfield  {author} {\bibinfo {author} {\bibfnamefont {W.~P.}\ \bibnamefont
  {Schleich}},\ }\href@noop {} {\emph {\bibinfo {title} {Quantum Optics in
  Phase Space}}}\ (\bibinfo  {publisher} {Wiley-VCH},\ \bibinfo {year}
  {2001})\BibitemShut {NoStop}%
\bibitem [{\citenamefont {Imre}\ \emph {et~al.}(1967)\citenamefont {Imre},
  \citenamefont {Ozizmir}, \citenamefont {Rosenbaum},\ and\ \citenamefont
  {Zweifel}}]{Imre1967}%
  \BibitemOpen
  \bibfield  {author} {\bibinfo {author} {\bibfnamefont {K.}~\bibnamefont
  {Imre}}, \bibinfo {author} {\bibfnamefont {E.}~\bibnamefont {Ozizmir}},
  \bibinfo {author} {\bibfnamefont {M.}~\bibnamefont {Rosenbaum}}, \ and\
  \bibinfo {author} {\bibfnamefont {P.~F.}\ \bibnamefont {Zweifel}},\ }\href
  {\doibase 10.1063/1.1705323} {\bibfield  {journal} {\bibinfo  {journal} {J.
  Math. Phys}\ }\textbf {\bibinfo {volume} {8}},\ \bibinfo {pages} {1097}
  (\bibinfo {year} {1967})}\BibitemShut {NoStop}%
\bibitem [{\citenamefont {Talkner}\ \emph {et~al.}(2008)\citenamefont
  {Talkner}, \citenamefont {Burada},\ and\ \citenamefont
  {H\"anggi}}]{Talkner2008}%
  \BibitemOpen
  \bibfield  {author} {\bibinfo {author} {\bibfnamefont {P.}~\bibnamefont
  {Talkner}}, \bibinfo {author} {\bibfnamefont {P.~S.}\ \bibnamefont {Burada}},
  \ and\ \bibinfo {author} {\bibfnamefont {P.}~\bibnamefont {H\"anggi}},\
  }\href {\doibase 10.1103/PhysRevE.78.011115} {\bibfield  {journal} {\bibinfo
  {journal} {Phys. Rev. E}\ }\textbf {\bibinfo {volume} {78}},\ \bibinfo
  {pages} {011115} (\bibinfo {year} {2008})}\BibitemShut {NoStop}%
\bibitem [{\citenamefont {Deffner}\ and\ \citenamefont
  {Lutz}(2008)}]{Deffner2008}%
  \BibitemOpen
  \bibfield  {author} {\bibinfo {author} {\bibfnamefont {S.}~\bibnamefont
  {Deffner}}\ and\ \bibinfo {author} {\bibfnamefont {E.}~\bibnamefont {Lutz}},\
  }\href {\doibase 10.1103/PhysRevE.77.021128} {\bibfield  {journal} {\bibinfo
  {journal} {Phys. Rev. E}\ }\textbf {\bibinfo {volume} {77}},\ \bibinfo
  {pages} {021128} (\bibinfo {year} {2008})}\BibitemShut {NoStop}%
\bibitem [{Note1()}]{Note1}%
  \BibitemOpen
  \bibinfo {note} {Now its mass and frequency have time
  parameters.}\BibitemShut {Stop}%
\bibitem [{\citenamefont {Landau}\ and\ \citenamefont
  {Lifshitz}(1976)}]{Landau1976}%
  \BibitemOpen
  \bibfield  {author} {\bibinfo {author} {\bibfnamefont {L.~D.}\ \bibnamefont
  {Landau}}\ and\ \bibinfo {author} {\bibfnamefont {E.~M.}\ \bibnamefont
  {Lifshitz}},\ }\href@noop {} {\emph {\bibinfo {title} {Mechanics}}},\
  \bibinfo {edition} {third edition}\ ed.\ (\bibinfo  {publisher} {Pergamon},\
  \bibinfo {year} {1976})\BibitemShut {NoStop}%
\bibitem [{\citenamefont {Goldstein}(1980)}]{goldstein1980}%
  \BibitemOpen
  \bibfield  {author} {\bibinfo {author} {\bibfnamefont {H.}~\bibnamefont
  {Goldstein}},\ }\href@noop {} {\emph {\bibinfo {title} {Classical
  Mechanics}}}\ (\bibinfo  {publisher} {Addison-Wesley},\ \bibinfo {year}
  {1980})\BibitemShut {NoStop}%
\bibitem [{\citenamefont {Funo}\ and\ \citenamefont
  {Quan}(2018{\natexlab{a}})}]{Ken2018a}%
  \BibitemOpen
  \bibfield  {author} {\bibinfo {author} {\bibfnamefont {K.}~\bibnamefont
  {Funo}}\ and\ \bibinfo {author} {\bibfnamefont {H.~T.}\ \bibnamefont
  {Quan}},\ }\href {\doibase 10.1103/PhysRevLett.121.040602} {\bibfield
  {journal} {\bibinfo  {journal} {Phys. Rev. Lett.}\ }\textbf {\bibinfo
  {volume} {121}},\ \bibinfo {pages} {040602} (\bibinfo {year}
  {2018}{\natexlab{a}})}\BibitemShut {NoStop}%
\bibitem [{\citenamefont {Funo}\ and\ \citenamefont
  {Quan}(2018{\natexlab{b}})}]{Ken2018b}%
  \BibitemOpen
  \bibfield  {author} {\bibinfo {author} {\bibfnamefont {K.}~\bibnamefont
  {Funo}}\ and\ \bibinfo {author} {\bibfnamefont {H.~T.}\ \bibnamefont
  {Quan}},\ }\href {\doibase 10.1103/PhysRevE.98.012113} {\bibfield  {journal}
  {\bibinfo  {journal} {Phys. Rev. E}\ }\textbf {\bibinfo {volume} {98}},\
  \bibinfo {pages} {012113} (\bibinfo {year} {2018}{\natexlab{b}})}\BibitemShut
  {NoStop}%
\bibitem [{\citenamefont {Campisi}(2008)}]{Campisi2008}%
  \BibitemOpen
  \bibfield  {author} {\bibinfo {author} {\bibfnamefont {M.}~\bibnamefont
  {Campisi}},\ }\href {\doibase 10.1103/PhysRevE.78.051123} {\bibfield
  {journal} {\bibinfo  {journal} {Phys. Rev. E}\ }\textbf {\bibinfo {volume}
  {78}},\ \bibinfo {pages} {051123} (\bibinfo {year} {2008})}\BibitemShut
  {NoStop}%
\bibitem [{\citenamefont {Atkins}\ and\ \citenamefont
  {Friedman}(1999)}]{Atkins1999}%
  \BibitemOpen
  \bibfield  {author} {\bibinfo {author} {\bibfnamefont {P.~W.}\ \bibnamefont
  {Atkins}}\ and\ \bibinfo {author} {\bibfnamefont {R.~S.}\ \bibnamefont
  {Friedman}},\ }\href@noop {} {\emph {\bibinfo {title} {Molecular Quantum
  Mechanics}}},\ \bibinfo {edition} {3rd}\ ed.\ (\bibinfo  {publisher} {Oxford
  University Press},\ \bibinfo {year} {1999})\BibitemShut {NoStop}%
\bibitem [{\citenamefont {Hutchisson}(1930)}]{Hutchisson1930}%
  \BibitemOpen
  \bibfield  {author} {\bibinfo {author} {\bibfnamefont {E.}~\bibnamefont
  {Hutchisson}},\ }\href {\doibase 10.1103/PhysRev.36.410} {\bibfield
  {journal} {\bibinfo  {journal} {Phys. Rev.}\ }\textbf {\bibinfo {volume}
  {36}},\ \bibinfo {pages} {410} (\bibinfo {year} {1930})}\BibitemShut
  {NoStop}%
\bibitem [{Note2()}]{Note2}%
  \BibitemOpen
  \bibinfo {note} {This result has been reobtained by many authors, e.g.,
  Refs.~\cite {Husimi1953,Dowling1991,Talkner2008}.}\BibitemShut {Stop}%
\bibitem [{\citenamefont {Wheeler}(1985)}]{Wheeler1985}%
  \BibitemOpen
  \bibfield  {author} {\bibinfo {author} {\bibfnamefont {J.~A.}\ \bibnamefont
  {Wheeler}},\ }\href@noop {} {\bibfield  {journal} {\bibinfo  {journal} {Lett.
  Math. Phys.}\ }\textbf {\bibinfo {volume} {10}},\ \bibinfo {pages} {201}
  (\bibinfo {year} {1985})}\BibitemShut {NoStop}%
\bibitem [{Note3()}]{Note3}%
  \BibitemOpen
  \bibinfo {note} {{\protect For the systems with degrees of freedom
  greater than one, the expressions of the ${\mathchar '26\mkern -9muh}^2$
  terms in Eqs.~(\ref {equationmotioninphasespacetillsecondorder2}) and~(\ref
  {expandedexponentialHamiltonian}) must be extended to account for the
  multidimensional case; see the corresponding terms in Eqs. (2.56) and (2.82)
  of Ref.~\cite {Hillery1984}. Although our approximation method is still
  applicable, because we face the extra numerical difficulty of solving
  multidimensional PDEs, we do not consider them in this paper. }}\BibitemShut
  {NoStop}%
\bibitem [{\citenamefont {Oppenheim}\ and\ \citenamefont
  {Ross}(1957)}]{Oppenheim1957}%
  \BibitemOpen
  \bibfield  {author} {\bibinfo {author} {\bibfnamefont {I.}~\bibnamefont
  {Oppenheim}}\ and\ \bibinfo {author} {\bibfnamefont {J.}~\bibnamefont
  {Ross}},\ }\href {\doibase 10.1103/PhysRev.107.28} {\bibfield  {journal}
  {\bibinfo  {journal} {Phys. Rev.}\ }\textbf {\bibinfo {volume} {107}},\
  \bibinfo {pages} {28} (\bibinfo {year} {1957})}\BibitemShut {NoStop}%
\bibitem [{\citenamefont {Zheng}\ and\ \citenamefont
  {Poletti}(2014)}]{Zheng2014}%
  \BibitemOpen
  \bibfield  {author} {\bibinfo {author} {\bibfnamefont {Y.}~\bibnamefont
  {Zheng}}\ and\ \bibinfo {author} {\bibfnamefont {D.}~\bibnamefont
  {Poletti}},\ }\href {\doibase 10.1103/PhysRevE.90.012145} {\bibfield
  {journal} {\bibinfo  {journal} {Phys. Rev. E}\ }\textbf {\bibinfo {volume}
  {90}},\ \bibinfo {pages} {012145} (\bibinfo {year} {2014})}\BibitemShut
  {NoStop}%
\bibitem [{\citenamefont {Berry}\ and\ \citenamefont
  {Klein}(1984)}]{Berry1984}%
  \BibitemOpen
  \bibfield  {author} {\bibinfo {author} {\bibfnamefont {M.~V.}\ \bibnamefont
  {Berry}}\ and\ \bibinfo {author} {\bibfnamefont {G.}~\bibnamefont {Klein}},\
  }\href@noop {} {\bibfield  {journal} {\bibinfo  {journal} {J. Phys. A: Math.
  Gen}\ }\textbf {\bibinfo {volume} {17}},\ \bibinfo {pages} {1805} (\bibinfo
  {year} {1984})}\BibitemShut {NoStop}%
\bibitem [{\citenamefont {Leonard}\ and\ \citenamefont
  {Deffner}(2015)}]{Leonard2015}%
  \BibitemOpen
  \bibfield  {author} {\bibinfo {author} {\bibfnamefont {A.}~\bibnamefont
  {Leonard}}\ and\ \bibinfo {author} {\bibfnamefont {S.}~\bibnamefont
  {Deffner}},\ }\href@noop {} {\bibfield  {journal} {\bibinfo  {journal} {Chem.
  Phys.}\ }\textbf {\bibinfo {volume} {446}},\ \bibinfo {pages} {18} (\bibinfo
  {year} {2015})}\BibitemShut {NoStop}%
\bibitem [{Note4()}]{Note4}%
  \BibitemOpen
  \bibinfo {note} {Now the mass and frequency are time dependent.}\BibitemShut
  {Stop}%
\bibitem [{Note5()}]{Note5}%
  \BibitemOpen
  \bibinfo {note} {If $i\eta \rightarrow -\beta $, then the equation is named
  the Bloch equation~\cite {Imre1967,Oppenheim1957}}\BibitemShut {NoStop}%
\bibitem [{\citenamefont {Dowling}\ \emph {et~al.}(1991)\citenamefont
  {Dowling}, \citenamefont {Schleich},\ and\ \citenamefont
  {Wheeler}}]{Dowling1991}%
  \BibitemOpen
  \bibfield  {author} {\bibinfo {author} {\bibfnamefont {J.~P.}\ \bibnamefont
  {Dowling}}, \bibinfo {author} {\bibfnamefont {W.~P.}\ \bibnamefont
  {Schleich}}, \ and\ \bibinfo {author} {\bibfnamefont {J.~A.}\ \bibnamefont
  {Wheeler}},\ }\href {\doibase 10.1002/andp.19915030702} {\bibfield  {journal}
  {\bibinfo  {journal} {Ann. der Phy.}\ }\textbf {\bibinfo {volume} {503}},\
  \bibinfo {pages} {423} (\bibinfo {year} {1991})}\BibitemShut {NoStop}%
\end{thebibliography}
%

\end{document}